\pdfoutput=1
\documentclass[preprint2]{emulateapj}
\usepackage{amsmath}
\usepackage{natbib}
\usepackage{graphicx}
\usepackage{epsf}
\usepackage{color}
\usepackage{threeparttable}
\usepackage{epsfig}
\DeclareGraphicsExtensions{.jpg,.pdf,.png,.eps,.ps}

\newcommand{\ltsima}{$\; \buildrel < \over \sim \;$}
\newcommand{\ltsim}{\lower.5ex\hbox{\ltsima}}

\newcommand{\five}{$5^\mathrm{h} 30^\mathrm{m}$}

\newcommand{\eg}{\textit{e.g.}}
\newcommand{\ie}{\textit{i.e.}}
\newcommand{\neff}{\ensuremath{N_\mathrm{eff}}}
\newcommand{\yhe}{\ensuremath{Y_p}}
\newcommand{\nrun}{\ensuremath{dn_s/d\ln k}}
\newcommand{\alens}{\ensuremath{A_{L}}}
\newcommand{\ns}{\ensuremath{n_{s}}}

\def\microKsq{\mu{\mbox{K}}^2}
\def \five {\textsc{ra5h30dec-55} }
\def \twthree {\textsc{ra23h30dec-55} }
\def \twonesix {\textsc{ra21hdec-60} }
\def \twonefif {\textsc{ra21hdec-50} }
\def \three {\textsc{ra3h30dec-60} }

\begin{document}

\title{A Measurement of the Damping Tail of the Cosmic Microwave Background Power Spectrum with the South Pole Telescope}

\author{
 R.~Keisler,\altaffilmark{1,2}
 C.~L.~Reichardt,\altaffilmark{3}
 K.~A.~Aird,\altaffilmark{4}
 B.~A.~Benson,\altaffilmark{1,5}
 L.~E.~Bleem,\altaffilmark{1,2}
 J.~E.~Carlstrom,\altaffilmark{1,2,5,6,7}
 C.~L.~Chang,\altaffilmark{1,5,7}
 H.~M. Cho, \altaffilmark{8}
 T.~M.~Crawford,\altaffilmark{1,6}
 A.~T.~Crites,\altaffilmark{1,6}
 T.~de~Haan,\altaffilmark{9}
 M.~A.~Dobbs,\altaffilmark{9}
 J.~Dudley,\altaffilmark{9}
 E.~M.~George,\altaffilmark{3}
 N.~W.~Halverson,\altaffilmark{10}
 G.~P.~Holder,\altaffilmark{9}
 W.~L.~Holzapfel,\altaffilmark{3}
 S.~Hoover,\altaffilmark{1,2}
 Z.~Hou,\altaffilmark{11}
 J.~D.~Hrubes,\altaffilmark{4}
 M.~Joy,\altaffilmark{12}
 L.~Knox,\altaffilmark{11}
 A.~T.~Lee,\altaffilmark{3,13}
 E.~M.~Leitch,\altaffilmark{1,6}
 M.~Lueker,\altaffilmark{14}
 D.~Luong-Van,\altaffilmark{4}
 J.~J.~McMahon,\altaffilmark{15}
 J.~Mehl,\altaffilmark{1}
 S.~S.~Meyer,\altaffilmark{1,2,5,6}
 M.~Millea,\altaffilmark{11}
 J.~J.~Mohr,\altaffilmark{16,17,18}
 T.~E.~Montroy,\altaffilmark{19}
 T.~Natoli,\altaffilmark{1,2}
 S.~Padin,\altaffilmark{1,6,14}
 T.~Plagge,\altaffilmark{1,6}
 C.~Pryke,\altaffilmark{1,5,6,20}
 J.~E.~Ruhl,\altaffilmark{19}
 K.~K.~Schaffer,\altaffilmark{1,5,21}
 L.~Shaw,\altaffilmark{22}
 E.~Shirokoff,\altaffilmark{3} 
 H.~G.~Spieler,\altaffilmark{13}
 Z.~Staniszewski,\altaffilmark{19}
 A.~A.~Stark,\altaffilmark{23}
 K.~Story,\altaffilmark{1,2}
 A.~van~Engelen,\altaffilmark{9}
 K.~Vanderlinde,\altaffilmark{9}
 J.~D.~Vieira,\altaffilmark{14}
 R.~Williamson,\altaffilmark{1,6} and
 O.~Zahn\altaffilmark{24}
}

\altaffiltext{1}{Kavli Institute for Cosmological Physics,
University of Chicago, 5640 South Ellis Avenue, Chicago, IL, USA 60637}
\altaffiltext{2}{Department of Physics,
University of Chicago,
5640 South Ellis Avenue, Chicago, IL, USA 60637}
\altaffiltext{3}{Department of Physics,
University of California, Berkeley, CA, USA 94720}
\altaffiltext{4}{University of Chicago,
5640 South Ellis Avenue, Chicago, IL, USA 60637}
\altaffiltext{5}{Enrico Fermi Institute,
University of Chicago,
5640 South Ellis Avenue, Chicago, IL, USA 60637}
\altaffiltext{6}{Department of Astronomy and Astrophysics,
University of Chicago,
5640 South Ellis Avenue, Chicago, IL, USA 60637}
\altaffiltext{7}{Argonne National Laboratory, 9700 S. Cass Avenue, Argonne, IL, USA 60439}
\altaffiltext{8}{NIST Quantum Devices Group, 325 Broadway Mailcode 817.03, Boulder, CO, USA 80305}
\altaffiltext{9}{Department of Physics,
McGill University, 3600 Rue University, 
Montreal, Quebec H3A 2T8, Canada}
\altaffiltext{10}{Department of Astrophysical and Planetary Sciences and Department of Physics,
University of Colorado,
Boulder, CO, USA 80309}
\altaffiltext{11}{Department of Physics, 
University of California, One Shields Avenue, Davis, CA, USA 95616}
\altaffiltext{12}{Department of Space Science, VP62,
NASA Marshall Space Flight Center,
Huntsville, AL, USA 35812}
\altaffiltext{13}{Physics Division,
Lawrence Berkeley National Laboratory,
Berkeley, CA, USA 94720}
\altaffiltext{14}{California Institute of Technology, MS 249-17, 1216 E. California Blvd., Pasadena, CA, USA 91125}
\altaffiltext{15}{Department of Physics, University of Michigan, 450 Church Street, Ann  
Arbor, MI, USA 48109}
\altaffiltext{16}{Department of Physics,
Ludwig-Maximilians-Universit\"{a}t,
Scheinerstr.\ 1, 81679 M\"{u}nchen, Germany}
\altaffiltext{17}{Excellence Cluster Universe,
Boltzmannstr.\ 2, 85748 Garching, Germany}
\altaffiltext{18}{Max-Planck-Institut f\"{u}r extraterrestrische Physik,
Giessenbachstr.\ 85748 Garching, Germany}
\altaffiltext{19}{Physics Department, Center for Education and Research in Cosmology 
and Astrophysics, 
Case Western Reserve University,
Cleveland, OH, USA 44106}
\altaffiltext{20}{Department of Physics, University of Minnesota, 116 Church Street S.E. Minneapolis, MN, USA 55455}
\altaffiltext{21}{Liberal Arts Department, 
School of the Art Institute of Chicago, 
112 S Michigan Ave, Chicago, IL, USA 60603}
\altaffiltext{22}{Department of Physics, Yale University, P.O. Box 208210, New Haven,
CT, USA 06520-8120}
\altaffiltext{23}{Harvard-Smithsonian Center for Astrophysics,
60 Garden Street, Cambridge, MA, USA 02138}
\altaffiltext{24}{Berkeley Center for Cosmological Physics,
Department of Physics, University of California, and Lawrence Berkeley
National Labs, Berkeley, CA, USA 94720}

\email{rkeisler@uchicago.edu}

\begin{abstract}

We present a measurement of the angular power spectrum of the cosmic microwave background (CMB) using data from the South Pole Telescope (SPT).  The data consist of 790 square degrees of sky observed at $150$~GHz during 2008 and 2009.  Here we present the power spectrum over the multipole range $650 < \ell < 3000$, where it is dominated by primary CMB anisotropy.  We combine this power spectrum with the power spectra from the seven-year Wilkinson Microwave Anisotropy Probe (WMAP) data release to constrain cosmological models.  We find that the SPT and WMAP data are consistent with each other and, when combined, are well fit by a spatially flat, $\Lambda$CDM cosmological model.  The SPT+WMAP constraint on the spectral index of scalar fluctuations is $n_s = 0.9663 \pm 0.0112$.  We detect, at $\sim$5$\sigma$ significance, the effect of gravitational lensing on the CMB power spectrum, and find its amplitude to be consistent with the $\Lambda$CDM cosmological model.  We explore a number of extensions beyond the $\Lambda$CDM model.  Each extension is tested independently, although there are degeneracies between some of the extension parameters.  We constrain the tensor-to-scalar ratio to be $r<0.21$ (95\% CL) and constrain the running of the scalar spectral index to be $\nrun=-0.024 \pm 0.013$.  We strongly detect the effects of primordial helium and neutrinos on the CMB; a model without helium is rejected at 7.7$\sigma$, while a model without neutrinos is rejected at 7.5$\sigma$.  The primordial helium abundance is measured to be $\yhe=0.296 \pm 0.030$, and the effective number of relativistic species is measured to be $\neff=3.85 \pm 0.62$.  The constraints on these models are strengthened when the CMB data are combined with measurements of the Hubble constant and the baryon acoustic oscillation feature.  Notable improvements include $n_s = 0.9668\pm0.0093$, $r<0.17$ (95\% CL), and $\neff=3.86 \pm 0.42$.  The SPT+WMAP data show a mild preference for low power in the CMB damping tail, and while this preference may be accommodated by models that have a negative spectral running, a high primordial helium abundance, or a high effective number of relativistic species, such models are disfavored by the abundance of low-redshift galaxy clusters.
  \end{abstract}


\bigskip\bigskip


\section{Introduction}
\label{sec:intro}

Measurements of anisotropy in the temperature of the cosmic microwave background (CMB) are among the most informative and robust probes of cosmology.  The acoustic oscillations of the primordial plasma have been measured on degree scales ($\ell \lesssim 500$) with cosmic-variance-limited precision by the Wilkinson Microwave Anisotropy Probe (WMAP) \citep{larson10}, yielding a wealth of cosmological information \citep{komatsu11}.  On much smaller scales, $\ell>3000$, the millimeter-wave anisotropy is dominated by secondary anisotropies from the Sunyaev-Zel'dovich (SZ) effects and by emission from foreground galaxies.  The thermal SZ effect arises from the scattering of CMB photons off the hot gas in gravitationally collapsed structures \citep{sunyaev72}, and thereby encodes information on the amplitude of matter fluctuations at intermediate redshifts.  Recently, \citet{lueker10} reported the first statistical measurement of the SZ effect using multi-frequency South Pole Telescope (SPT) data.  This was followed by a measurement using data from the Atacama Cosmology Telescope \citep[ACT,][]{das10, dunkley10} and an improved SPT measurement \citep{shirokoff11}.  The angular power spectrum of millimeter-wave emission from high-redshift, dusty, star-forming galaxies has also been characterized by SPT, ACT, and Planck \citep{hall10, dunkley10, shirokoff11, planck11-6.6_arxiv}.

On intermediate scales, $500 < \ell < 3000$, the primary CMB anisotropy is the dominant source of millimeter-wave anisotropy, but its power is falling exponentially with decreasing angular scale.  The reduction in CMB power is due to the diffusion of photons in the primordial plasma and is often referred to as Silk damping \citep{silk68}.  This ``damping tail'' of the primary CMB anisotropy has been measured by a number of experiments, notably the Arcminute Cosmology Bolometer Array Receiver \citep[ACBAR,][]{reichardt09a}, QUEST at DASI \citep[QUaD,][]{brown09, friedman09}, and ACT \citep{das10}.

Measurements of the CMB damping tail, in conjunction with WMAP's measurements of the degree-scale CMB anisotropy, provide a powerful probe of early-universe physics. The damping tail measurements significantly increase the angular dynamic range of CMB measurements and thereby improve the constraints on inflationary parameters such as the scalar spectral index and the amplitude of tensor fluctuations.  Measurements of the angular scale of the damping can constrain the primordial helium abundance and the effective number of relativistic particle species during the radiation-dominated era.  Finally, the damping tail is altered at the few-percent level by gravitational lensing of the CMB, and is therefore sensitive to the matter fluctuations at intermediate redshifts.

The work presented here is a measurement of the CMB damping tail using data from the SPT.  The data were taken at $150$~GHz during 2008 and 2009 and cover approximately 790 square degrees of sky.  This is approximately four times the area used in the preceding SPT power spectrum result, \citet{shirokoff11}.  The new power spectrum spans the multipole range $650 < \ell < 3000$ (angular scales of approximately  $4^\prime < \theta < 16^\prime$) and is dominated by primary CMB temperature anisotropy.

The paper is organized as follows.  We describe the SPT, the observations used in this analysis, and the pipeline used to process the raw data into calibrated maps in Section~\ref{sec:obs_and_red}.  We discuss the pipeline used to process the maps into an angular power spectrum in Section~\ref{sec:powspec}.  We combine the SPT power spectrum with external data, most importantly the seven-year WMAP data release, to constrain cosmological models in Section~\ref{sec:cosmo}, and we conclude in Section~\ref{sec:conclusion}.


\section{Observations and Data Reduction}
\label{sec:obs_and_red}

The SPT is a 10-meter diameter off-axis Gregorian telescope located at the South Pole.  The current receiver is equipped with 960 horn-coupled spiderweb bolometers with superconducting transition-edge sensors.  The receiver included science-quality detectors at frequency bands centered at approximately $150$ and $220$~GHz in 2008, and at $95$, $150$, and $220$~GHz in 2009.  The telescope and receiver are discussed in further detail in \citet{ruhl04}, \citet{padin08}, and \citet{carlstrom11}.

\subsection{Fields and Observation Strategy}
\label{sec:fields_and_obs}
\begin{figure}[]\centering
\includegraphics[width=0.5\textwidth]{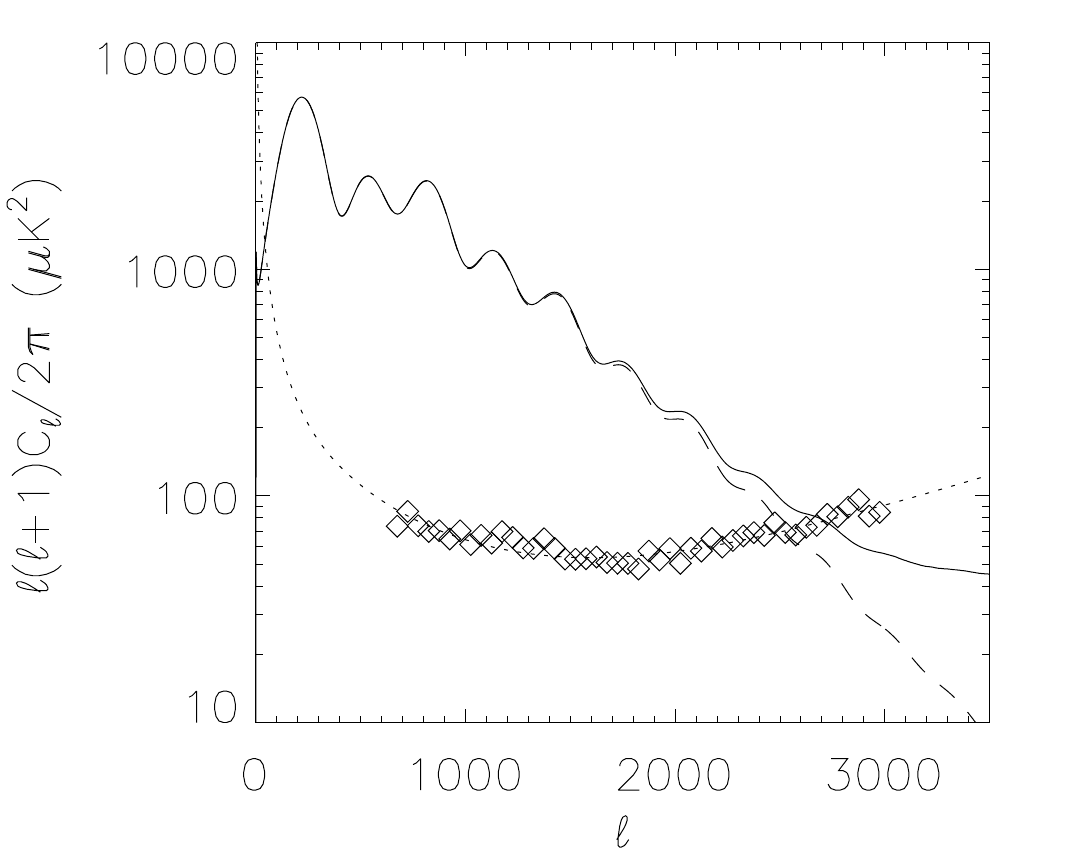}
  \caption[]{ The beam-deconvolved noise power in the SPT maps used in this analysis ({\it symbols} show data and {\it dotted line} shows two-component model) compared to theoretical power spectra including CMB only ({\it dashed line}) and CMB+foregrounds ({\it solid line}).  The precision of the power spectrum measurement is limited by sample variance rather than detector or atmospheric noise across most of the $650 < \ell < 3000$ range.}
  \label{fig:noise_power}
\end{figure}

In this work we use data at $150$~GHz taken during the 2008 and 2009 austral winters.  This includes five fields whose locations, shapes, and effective areas (\ie\ the area of the masks used in the power spectrum analysis) are given in Table~\ref{tab:fields}.  The total effective area is approximately 790 square degrees.\footnote{The \twonefif and \twonesix fields overlap slightly.  This reduces the effective total area of the power spectrum analysis.  We have ignored this effect in our simulations, and have therefore underestimated the SPT bandpower errors by at most 0.4\%.}  The mean beam-convolved noise power in these fields is approximated by the sum of a white noise component and a component that increases in power with decreasing $\ell$: $C_\ell = (17.9~\mu\rm{K}\rm{-arcmin})^2+3 \times 10^{-4}(\frac{\ell}{1000})^{-3.1} \microKsq$.\footnote{Throughout this work, the unit $\textrm{K}$ refers to equivalent fluctuations in the CMB temperature, i.e.,~the temperature fluctuation of a 2.73$\,$K blackbody that would be required to produce the same power fluctuation.  The conversion factor is given by the derivative of the blackbody spectrum, $\frac{dB}{dT}$, evaluated at 2.73$\,$K.}  The beam-deconvolved noise power and its two-component fit are shown in Figure~\ref{fig:noise_power}.

\begin{table*}[]
\centering
\caption[SPT Fields from 2008 and 2009]{SPT Fields from 2008 and 2009} \small
\begin{tabular}{ | l  c c c c c | }
\hline
Name & R.A. ($^\circ$) & Decl. ($^\circ$) & $\Delta$R.A. ($^\circ$) & $\Delta$Decl. ($^\circ$) & Effective Area (sq. degrees)\\
\hline \hline
\five & 82.5 & -55 & 15.7 & 10.3 &     91.6 \\
\twthree & 352.5   & -55  & 18.2 & 10.1 &  105.5\\
\twonesix  & 315.0  & -60 &  30.5 & 10.5 &  156.9\\
\three  & 52.5 & -60   & 45.3 & 10.6 & 236.0\\
\twonefif  & 315.0 & -50 & 30.2 & 10.5 & 202.1\\
\hline
Total      & & & & &   792.1\\
\hline
\end{tabular}
\label{tab:fields}
\begin{tablenotes}
\item The locations and sizes of the fields observed by SPT in 2008 and 2009.  For each field we give the center of the field in Right Ascension (R.A.) and Declination (Decl.), the extent of the field in Right Ascension and Declination, and the effective field area.
\end{tablenotes}
\end{table*}

The fields were observed with two different types of scan strategies.
The scan strategy used for the \five\ field consisted of
constant-elevation scans across the field. 
After each scan back
and forth in azimuth across the field, the telescope stepped 0.125$^\circ$ in elevation.
We refer to a complete set of scans covering the entire field as an observation.

The remaining four fields were observed using a lead/trail scan strategy.  In this strategy each field was divided into two halves in Right Ascension, and the two halves were observed sequentially using constant-elevation scans.  Due to the Earth's rotation, both halves of the field are observed at the same range of azimuth angle.
This strategy allows for the possible removal of ground-synchronous signal, although we see no evidence for such a signal at the angular scales of interest and do not difference the lead and trail maps.

\subsection{Time-ordered Data to Maps}
\label{sec:reduction_and_maps}

Each SPT detector measures the sky brightness temperature plus noise, and records this measurement as the time-ordered data (TOD).  The TOD are recorded at 100~Hz, so we have information in the TOD on signals up to 50~Hz.  For a typical scan speed and elevation, 50~Hz corresponds to a mode oscillating along the scan direction at $\ell \sim 60,000$. This analysis, which only measures power at $\ell < 3000$, can benefit computationally from using a down-sampled version of the TOD. We down-sample by a factor of six. Prior to down-sampling, we low-pass filter the TOD at 7.5 Hz.  The combined effect of the filter and down-sampling is negligibly small ($< 0.1\%$ in power) on the scales of interest and we do not correct for it.

The TOD are further low-pass filtered at 5 Hz as a safeguard against high-frequency noise being aliased into the signal band.  The TOD are effectively high-pass filtered by the removal of a Legendre polynomial from the TOD of each detector on each scan.  The order of the polynomial ranges from 9 to 18 and is chosen to have approximately the same number of degrees of freedom (dof) per unit angular distance ($\sim$1.5 dof per degree).  The polynomial fit removes low-frequency instrumental and atmospheric noise.  Regions of sky within 5 arcminutes of point sources with fluxes $S_{\rm{150 GHz}}>\ 50\ \rm{mJy}$ are masked during the polynomial fits.

Correlated atmospheric noise remains in the TOD after the bandpass filtering.
We remove the correlated noise by subtracting the mean signal across each bolometer 
wedge\footnote{The SPT array consists of 6 wedge-shaped bolometer 
modules, each with 160 detectors.  Wedges are configured with a set of filters that determine their observing frequency 
(e.g. $95$, $150$ or $220$~GHz).} at each time sample.  This subtraction serves as an approximately isotropic high-pass filter.

\begin{figure*}[h]\centering
\includegraphics[width=1.0\textwidth]{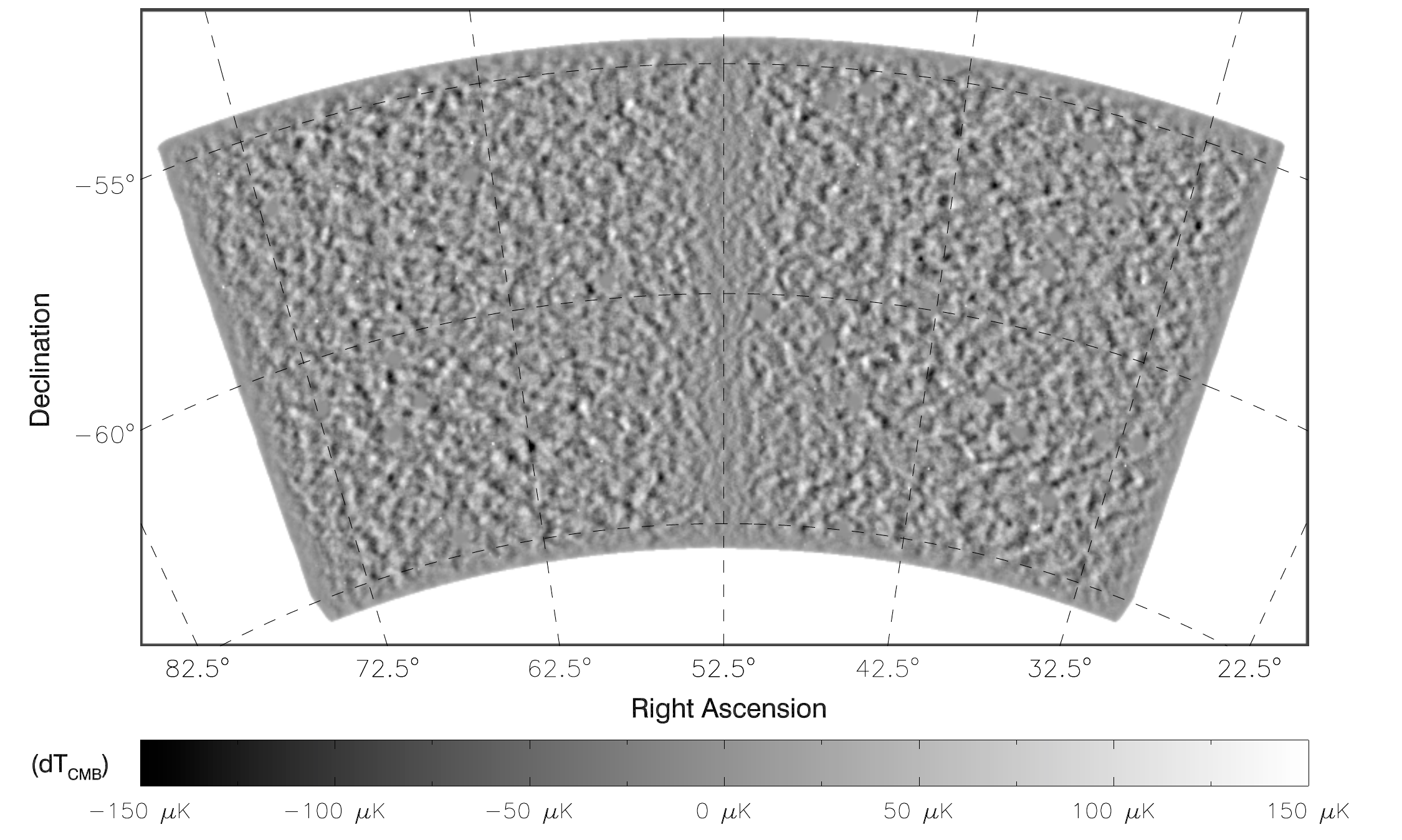}
  \caption[]{ A map of the \three field, which is typical of the fields used in this analysis.  The effective area is 236 square degrees.  The structure visible in this map is due to primary CMB anisotropy, not instrumental or atmospheric noise.  Modes with $\ell \lesssim 600$ are strongly suppressed due to the high-pass filtering of the time-ordered data.  The map has been multiplied by the apodization and point source masks described in Section~\ref{sec:windows}, such that bright point sources with $S_{\rm{150 GHz}}>\ 50\ \rm{mJy}$ have been masked.  A vertical stripe along the center of the map has been filtered more strongly than other regions.  This stripe lies on the boundary of the lead and trail fields and is caused by high-pass filtering the time-ordered data by removing polynomial functions.  This effect is accounted for in our analysis by using simulated observations.}
  \label{fig:3h_map}
\end{figure*}

The data from each detector receives a weight based on the power spectral density of its calibrated TOD in the 1-3 Hz band, which corresponds approximately to the signal band of the power spectrum analysis presented in Section~\ref{sec:powspec}.  We bin the data into map pixels based on the telescope pointing information.  The maps use the oblique Lambert equal-area azimuthal projection \citep{snyder87} with $1^\prime$ pixel resolution.  The power spectrum analysis presented in Section~\ref{sec:powspec} adopts the flat-sky approximation, for which wavenumber $k$ is equivalent to multipole $\ell$ and Fourier transforms replace spherical harmonic transforms.  The $150$~GHz map of the \three field is shown in Figure~\ref{fig:3h_map}.

\subsection{Beam Functions}
\label{sec:beams}
The optical response as a function of angle---or beam---of the SPT must be measured accurately in order to calibrate the signals in the maps as a function of angular scale.  Due to the limited dynamic range of the detectors, the SPT beams were measured by combining maps of three types of sources: Jupiter, Venus, and the five brightest point sources in the CMB fields.
The beam within a radius of $4^\prime$ is measured on the five brightest point sources in these fields, and this naturally takes into account the enlargement of the effective beam due to random errors in the pointing reconstruction.  
Outside of this $4^\prime$ radius, maps of Jupiter are used to constrain a diffuse, low-level sidelobe that accounts for roughly 15\% of the total beam solid angle.
Maps of Venus are used to stitch together the outer and inner beams.

\begin{figure}[h]\centering
\includegraphics[width=0.5\textwidth]{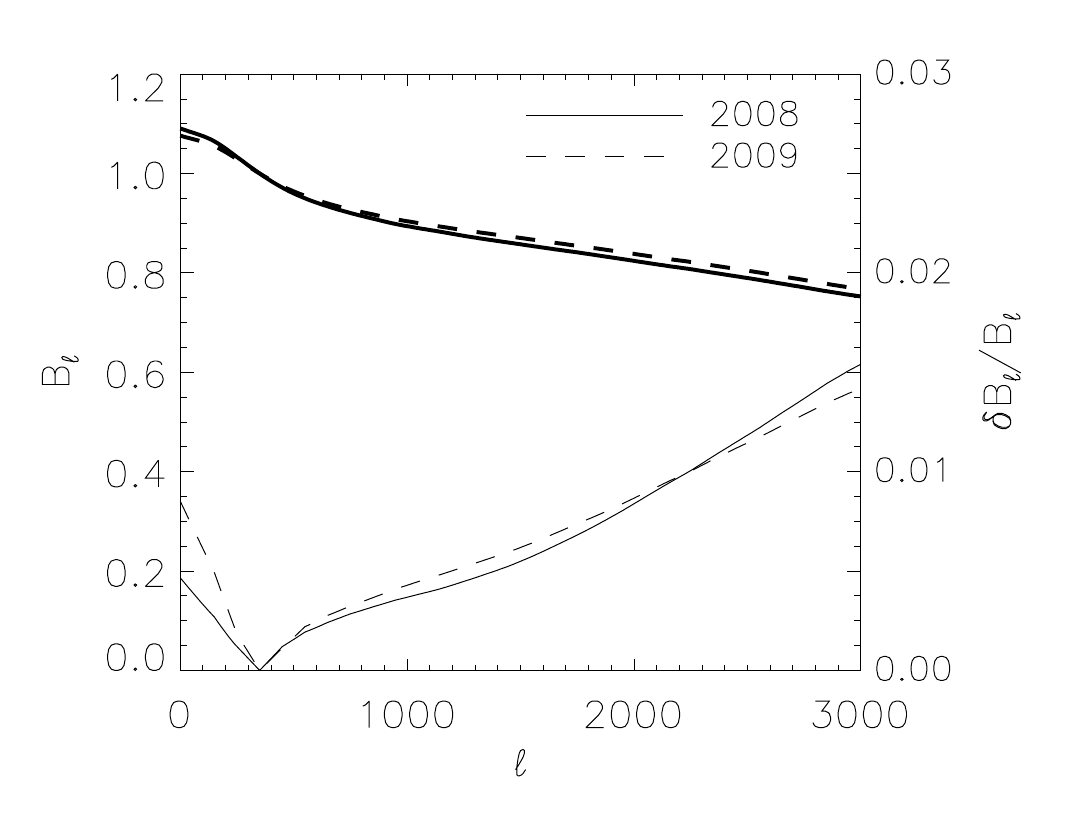}
  \caption[]{The $150$~GHz beam functions (bold, left scale) and fractional errors (thin, right scale).  The beam function is normalized to one at $\ell=350$.}
  \label{fig:beams}
\end{figure}

We measure the beam function, $B_\ell$, which is the azimuthally averaged Fourier transform of the beam map.  The beam function is normalized to unity at $\ell=350$.  We express our uncertainty in the beam as an uncertainty in $B_\ell$.  We account for the uncertainty arising from several statistical and systematic
effects, including residual atmospheric noise in the maps of Venus and
Jupiter, and the weak dependence of $B_\ell$ on the choice of radius
used to stitch together the inner and outer beam maps.  The different sources of uncertainty are incorporated into the power spectrum analysis through the bin-to-bin covariance matrix, as described in Section~\ref{sec:cov}.  In Figure~\ref{fig:beams}, we show the beam functions and the quadrature sum of the different beam uncertainties, which
gives a sense of the total uncertainty.  There are small variations in the beam function between 2008 and 2009 across this multipole range due primarily to changes in the locations of the 150 GHz detector modules in the focal plane.  The beam function is uncertain at the percent level across the multipoles of the power spectrum presented here.  The SPT beams are discussed in more detail in \citet{lueker10} and Schaffer et al. (submitted).

\subsection{Calibration}
\label{sec:calibration}

The TOD are initially calibrated using a galactic HII region, RCW38.  The final calibration used in this analysis is calculated by comparing the SPT power spectrum described in Section~\ref{sec:powspec} with the seven-year WMAP temperature power spectrum reported in \citet{larson10}.  The power spectra are compared across the angular range $650 < \ell \,< 1000$, where the primary CMB anisotropy is the dominant source of power.  First we construct a binned WMAP spectrum $D_\ell^{\rm{WMAP}}$  that may be directly compared to the SPT spectrum by multiplying the $\delta\ell=1$ WMAP power spectrum by the SPT bandpower window functions described in Section~\ref{sec:bpwf}.  Next we calculate the ratio $R=D_\ell^{\rm{WMAP}}/D_\ell^{\rm{SPT}}$ and its weighted mean across this multipole range, $\langle R \rangle=\sum_i (w_i R_i) / \sum_i w_i$, where $w_i$ is the weight of the $i^{\rm{th}}$ bin and is calculated as $w = (D_\ell^{\rm{WMAP}})^2/(\sigma_{D_\ell^{\rm{WMAP}}}^2 + \sigma_{D_\ell^{\rm{SPT}}}^2)$.  The WMAP error $\sigma_{D_\ell^{\rm{WMAP}}}$ comes from the binned power spectrum provided by the WMAP team, which is also binned as $\delta\ell=50$ at these multipoles.

This calibration method is model-independent; it assumes only that the power measured in the SPT fields is statistically representative of the all-sky power measured by WMAP.  As such, this calibration method does not bias the constraints on cosmological parameters described in Section~\ref{sec:cosmo}.  We estimate the uncertainty in the SPT calibration to be 3.1\% in power.  This uncertainty is included in the bin-to-bin covariance matrix as described in Section~\ref{sec:combining}.


\section{Power Spectrum Analysis}
\label{sec:powspec}
In this section we describe the pipeline used to process the maps into an angular power spectrum.  The method closely follows the approach used by \cite{lueker10} and \cite{shirokoff11}.  We adopt the flat-sky approximation, in which the angular wavenumber $k$ is equivalent to multipole $\ell$ and Fourier transforms replace spherical harmonic transforms.    The distortion to the power spectrum due to adopting the flat-sky approximation on the SPT maps is negligibly small.  We refer to the power in a given band of angular frequencies as the \emph{bandpower}.  We report bandpowers in terms of $D_\ell$, where

\begin{equation}
D_\ell=\frac{\ell\left(\ell+1\right)}{2\pi} C_\ell\;.
\end{equation}

\subsection{Maps}
The power spectrum analysis begins with a set of $150$~GHz maps for each field.  Each map corresponds to a single bottom-to-top observation of the field.  For fields that were observed using a lead/trail method, we define a map to be a combination of consecutive lead and trail maps.  We do not subtract the lead and trail maps.  For the \twthree field, which was observed using large elevation steps, we take this one step further and define a map as the combination of a pair of consecutive lead/trail pairs.  The composite map has more uniform coverage than the individual maps due to small elevation offsets between the individual maps.

\subsection{Windows\label{sec:windows}}

The next step is to calculate the Fourier transform of each map, $\tilde{m}^{A}$, where $A$ is the observation index.  All maps of the same field are multiplied by the same window $\textbf{W}$ prior to the Fourier transform.  This window is the product of an apodization mask used to avoid sharp edges at the map borders and a point source mask used to reduce the power from bright point sources.  The apodization mask is a smoothed version of a map of zeros and ones in which a one denotes a pixel that was observed by at least one detector in every observation.  We mask all point sources that we measure to have $150$~GHz flux $> 50$ mJy.  Each point source is masked by a $5'$-radius disk, with a Gaussian taper outside this radius with $\sigma_{taper}=5'$.  This relatively broad Gaussian taper was chosen to minimize the mode-coupling due to the point source mask.  Using previous measurements of the mm-wave point source population \citep{vieira10, shirokoff11}, we estimate that the power from residual point sources below this flux cut is $C_\ell \sim 1.3 \times 10^{-5} \microKsq$, or $D_\ell \sim 18~\microKsq \left(\frac{\ell}{3000}\right)^2$.  This is approximately equal to the power from the primary CMB anisotropy at $\ell=3000$, the upper edge of the multipole range of this analysis.  A more aggressive point source cut, say 10 mJy, could have been used to further reduce the residual power, but the gains were not considered worth the cost of decreased sky area and increased mode-coupling.

\subsection{Cross-Spectra \label{sec:cross-spectra}}
The next step is to cross-correlate maps from different observations of the same field.  The noise in each individual observation map is assumed to be uncorrelated with the noise in all other maps, so the resulting cross-spectra are free from noise bias.  We calculate the cross-spectrum between maps from two different observations, $A$ and $B$, and average within $\ell$-bins:

\begin{equation}
\label{eqn:ddef}
 \widehat{D}^{AB}_b\equiv \left< \frac{\ell(\ell+1)}{2\pi}H_{\pmb{\ell}}Re[\tilde{m}^{A}_{\pmb{\ell}}\tilde{m}^{B*}_{\pmb{\ell}}] \right>_{\ell \in b}, 
\end{equation} where $H_{\pmb{\ell}}$ is a two-dimensional weight array described below and ${\pmb{\ell}}$ is a vector in the two-dimensional, gridded Fourier plane of $\delta\ell=10$ resolution.  We average $\widehat{D}^{AB}_b$ among all pairs of observations $A$ and $B$, where $A \ne B$, to produce $\widehat{D}_b$.  Each observation receives the same weight.  For a typical field there are approximately 200 maps and 20,000 cross-spectra.  The maps are zero-padded prior to the Fourier transform such that the native $\ell$ resolution is $\delta\ell=10$, which allows for clean separation into the final bins, which have a width of $\delta\ell=50$.  The lower edge of the lowest bin is $\ell=650$, while the upper edge of the upper bin is $\ell=3000$.

The noise in the maps used in this analysis is statistically anisotropic; for a fixed $\ell$, modes that oscillate perpendicular to the scan direction (here defined as $\ell_x=0$) are the most noisy.  For this reason, the modes that contribute to a given $\ell$-bin do not necessarily have uniform noise properties.  We construct a two-dimensional weight array to optimally combine the modes contributing to each $\ell$-bin, 

\begin{equation}
\label{eqn:2dweight}
H_{\pmb{\ell}} \propto (C_\ell^{\rm{th}} + N_{\pmb{\ell}})^{-2} \,,
\end{equation} where $C_\ell^{\rm{th}}$ is the theoretical power spectrum used in the simulations described in Section~\ref{sec:transfer} and $N_{\pmb{\ell}}$ is the two-dimensional, calibrated, beam-deconvolved noise power.  We smooth $(C_\ell^{\rm{th}} + N_{\pmb{\ell}})$ with a Gaussian kernel of width $\sigma_\ell$=425 in order to capture only the broad anisotropy of the noise power.  The weight array is normalized such that $\sum_{\ell \in b}H_{\pmb{\ell}}=1$ for each bin $b$.  The weights are approximately uniform for the sample-variance-dominated bins ($\ell<2700$) and begin to preferentially de-weight $\ell_x\simeq0$ modes for bins at $\ell>2700$.  We estimate that this weighting scheme reduces our bandpower errors by approximately 8\% in the highest bin.

\subsection{Jackknives}
\label{sec:jackknives}
Before proceeding with the rest of the power spectrum analysis, we apply a set of ``jackknife'' tests to the bandpowers to search for possible systematic errors. In a jackknife test, the data are divided into two halves associated with potential sources of systematic error. The two halves are differenced to remove any astronomical
signal, and the resulting power spectrum is compared to an ``expectation spectrum'', the power we expect to see in each jackknife spectrum in the absence of a systematic problem or spurious signal.  This expectation spectrum can be non-zero due to mundane observational effects (\eg\ a mid-season adjustment in the scanning strategy, or unequal weighting of left- and right-going scans).  We use simulations to estimate the expectation spectra and find that they are small, with $D_\ell<1\ \microKsq$ at all multipoles.  Significant deviations of the jackknife spectrum from the expectation spectrum could indicate either a systematic contamination of the data or a misestimate of the noise.  We construct difference maps (a single map in one jackknife half subtracted from a single map in the other half) and measure the jackknife spectrum as the average cross-spectrum between the difference maps, in a method analogous to that described in Section~\ref{sec:cross-spectra}.  The jackknife spectra are calculated for five broad $\ell$-bins.  We perform four jackknife tests based on the following criteria.

\begin{itemize}

\item Time: We split the data into the first and second halves of observation.  This tests for any systematic that might be changing on weekly or monthly timescales.

\item Scan Direction: We split the data into left-going and right-going halves.  This tests for scan-synchronous signals or any signal that is not time-symmetric, such as inaccurate deconvolution of the detector transfer functions. 

\item Azimuthal Range: We split the data into observations taken at azimuths that we expect to be more or less susceptible to ground pickup.  We determine these azimuths by making maps using ``ground-centered'' coordinates (Azimuth/Elevation) as opposed to ``sky-centered'' (R.A./Dec.).  Although we detect emission from the ground on large scales ($\ell\sim50$) in these ground-centered maps, we do not expect such emission to bias our measurement of the sky power, as our observations are spread randomly in azimuth.  We use the azimuth-based jackknife to test this assertion.

\item Moon: We split the data into observations taken at times when the Moon was either above or below the horizon.  This tests for any significant coupling to the moon via far sidelobes of the SPT beam.

\end{itemize}

For each test we calculate the $\chi^2$ of the jackknife spectrum with respect to the expectation model.  We calculate the probability to exceed (PTE) this $\chi^2$ for five degrees of freedom for each test and find PTE = 0.38, 0.05, 0.92, 0.31 for the Time, Scan Direction, Azimuthal Range, and Moon tests, respectively.  Although there are individual tests which have moderately low (0.05) or high (0.92) PTE's, they are consistent with a uniform distribution when taken as a whole.  We therefore find no significant evidence for systematic contamination of the SPT bandpowers.

\subsection{Unbiased Spectra}
\label{sec:unbiased}

The spectra calculated in Section~\ref{sec:cross-spectra} are biased estimates of the true sky power.  The unbiased spectra are

\begin{equation}
D_b\equiv \left(K^{-1}\right)_{bb^\prime}\widehat{D}_{b^\prime}
\end{equation} where $b^\prime$ is summed over.  The $K$ matrix accounts for the effects of the beams, TOD filtering, pixelization, windowing, and band-averaging. It can be expanded as

\begin{equation}
\label{eqn:kdef}
K_{bb^\prime}=P_{b\ell}\left(M_{\ell\ell^\prime}[\textbf{W}]\,F_{\ell^\prime}B^{2}_{\ell^\prime}\right)Q_{\ell^\prime b^\prime}
\end{equation} where $\ell$ and $\ell^\prime$ are summed over.  $Q_{\ell b}$ and $P_{b\ell}$ are the binning and re-binning operators \citep{hivon02}. $B_{\ell}$ is the beam function described in Section~\ref{sec:beams}.  $F_{\ell}$ is the transfer function due to TOD filtering and map pixelization, and is described in Section~\ref{sec:transfer}.  The ``mode-coupling matrix'' $M_{\ell\ell^\prime}[\textbf{W}]$ is due to observing a limited portion of the sky and is calculated analytically from the known window $\textbf{W}$, as described in \citet{lueker10}.  At the multipoles considered in this work, the elements of the mode-coupling kernel depend only on the distance from the diagonal.

\subsubsection{Transfer Function \label{sec:transfer}}
The transfer function $F_\ell$ is calculated from simulated observations of 1500 sky realizations (300 per field) that have been smoothed by the appropriate beam.  These simulations are also used to calculate the sample variance described in Section~\ref{sec:cov}, and it is therefore important that the power spectrum used to generate the simulated skies be consistent with previous measurements and with the power spectrum measured in this work.  The simulated skies are Gaussian realizations of the primary CMB from the best-fit lensed WMAP+CMB $\Lambda$CDM model from the seven-year WMAP release\footnote{http://lambda.gsfc.nasa.gov/} combined with contributions from randomly distributed point sources, clustered point sources, and the Sunyaev-Zel'dovich (SZ) effect.  The random point source component uses $C_\ell = 12.6\times10^{-6} \microKsq$ ($D_{\ell=3000} \equiv D_{3000}=18.1~\microKsq$).  The clustered point source component uses $D_\ell=3.5\ \microKsq f_\ell$, where $f_\ell=1$ for $\ell < 1500$ and $f_\ell = (\frac{\ell}{1500})^{0.8}$ for $\ell \ge 1500$.  This shape is designed to approximate the shape of the clustering power in both the linear and non-linear regimes \citep{shirokoff11, millea11}.  The SZ component uses the thermal SZ template of \cite{sehgal10}, which has a shape similar to the templates from more recent models \citep{trac11, shaw10, battaglia10}, normalized to $D_\ell=5.5\ \microKsq$ at $\ell=3000$.  The foreground components are consistent with the measurements of \cite{shirokoff11} and \cite{vieira10}, and the total power is consistent with the power spectrum presented in this analysis.

The simulated skies are observed using the SPT pointing information, filtered identically to the real data, and processed into maps.  The power spectrum of the simulated maps is compared to the known input spectrum to calculate the effective transfer function \citep{hivon02} using an iterative scheme.  The initial estimate is

\begin{equation}
\label{eqn:transguess}
F^{(0)}_{\ell}=\frac{\left<\widehat{D}_{\ell}\right>_\textrm{sim}}{w_2 {B_\ell}^2 D^{\textrm{th}}_\ell},
\end{equation} where the superscript $(0)$ indicates that this is the first iteration in the transfer function estimates.  We approximate the coupling matrix as diagonal for this initial estimate.  The factor $w_2= \int d\!x \textbf{W}^2$ is a normalization factor for the area of the window.  We then iterate on this estimate using the mode-coupling matrix:

\begin{equation}\label{eqn:transiter}
F^{(i+1)}_{\ell}=F^{(i)}_{\ell}+\frac{\left<\widehat{D}_{\ell}\right>_{\rm sim} -  M_{\ell\ell^\prime} F_{\ell^\prime}^{(i)} {B_{\ell^\prime}}^2 D^{\textrm{th}}_{\ell^\prime}}{w_2 {B_{\ell}}^2 D^{\textrm{th}}_{\ell} }
\end{equation} where $\ell^\prime$ is summed over.  The transfer function estimate has converged after the second iteration, and we use the fifth iteration.  The transfer function is equal to $\sim$$0.25$ at $\ell=650$ and plateaus to $\sim$$0.85$ at $\ell > 1200$; the transfer function does not reach unity because of the strong filtering of $\ell_x \lesssim 300$ modes.

\subsection{Bandpower Covariance Matrix}
\label{sec:cov}

The bandpower covariance matrix describes the bin-to-bin covariance of the unbiased spectrum and has signal term and a noise term.  The signal term is estimated using the bandpowers from the signal-only simulations described in Section~\ref{sec:transfer} and is referred to as the ``sample variance.''  The noise term is estimated directly from the data using the distribution of the cross-spectrum bandpowers $D^{AB}_b$ between observations A and B, as described in \citet{lueker10}, and is referred to as the ``noise variance.''  The covariance is dominated by sample variance at low multipoles and noise variance at high multipoles, with the two being equal at $\ell\sim2700$.

The initial estimate for the bandpower covariance matrix is poor for off-diagonal elements.  We expect some statistical uncertainty, 

\begin{equation}
\left<\left(\textbf{C}_{ij}-\left<\textbf{C}_{ij}\right>\right)^2\right>=\frac{\textbf{C}_{ij}^2+\textbf{C}_{ii}\textbf{C}_{jj}}{n_{obs}}.
\end{equation} This uncertainty is significantly higher than the true covariance for almost all off-diagonal terms due to its dependence on the large diagonal covariances.  We reduce the impact of this uncertainty by ``conditioning" the covariance matrix in the following manner.  First we introduce the correlation matrix

\begin{equation}
\pmb{\rho}_{ij} = \frac{\textbf{C}_{ij}}{\sqrt{\textbf{C}_{ii}\textbf{C}_{jj}}}.
\end{equation}

The shape of the correlation matrix is determined by the mode-coupling matrix and is a function only of the distance from the diagonal.  We calculate the conditioned correlation matrix by averaging the off-diagonal elements at a fixed separation from the diagonal:
\begin{equation}
\label{eqn:covcond}
\pmb{\rho}^\prime_{ii^\prime}=\frac{
\sum_{i_1-i_2=i-i^\prime}\pmb{\rho}_{i_1i_2}}{\sum_{i_1-i_2=i-i^\prime} 1}.
\end{equation}

We set $\pmb{\rho}^\prime_{ij}=0$ for all off-diagonal elements that are a distance $\ell>250$ from the diagonal.  The conditioned covariance matrix is then
\begin{equation}
\textbf{C}^\prime_{ij} = \pmb{\rho}^\prime_{ij}\sqrt{\textbf{C}_{ii}\textbf{C}_{jj}}.
\end{equation}

We must also consider the bin-to-bin covariance due to the uncertainties in the beam function $B_\ell$ as described in Section~\ref{sec:beams}.  We construct a ``beam correlation matrix'' for each of the sources of beam uncertainty described in Section~\ref{sec:beams}:

\begin{equation}
\pmb{\rho}^{beam}_{ij} = \left(\frac{\delta D_i}{D_i}\right) \left(\frac{\delta D_j}{D_j}\right)
\end{equation}
where
\begin{equation}
\frac{\delta D_i}{D_i} = 1-\left(1+\frac{\delta B_i}{B_i}\right)^{-2}.
\end{equation}

The combined beam correlation matrix is the sum of the beam correlation matrices due to each of the sources of uncertainty.  The beam covariance matrix is then

\begin{equation}
\textbf{C}^{beam}_{ij} = \pmb{\rho}^{beam}_{ij}D_{i}D_{j}.
\end{equation}

We calculate the beam covariance matrix for each year and add them to the bandpower covariance matrix as described in Section~\ref{sec:combining}.

\subsection{Combining Different Fields \label{sec:combining}}
We have five sets of bandpowers and covariances, one set per field, which must be combined into a single set of bandpowers and covariances.  In the limit that the noise properties of all fields are identical, or in the limit that the precision of the power spectrum is limited by sample variance on all scales of interest, each field would be weighted by its effective area (\ie\ the area of its window).  While neither of these conditions is exactly true for our fields (the \twonefif field has higher noise than the other fields, and the power spectrum is dominated by noise variance at $\ell>2700$), they are both nearly true, and we use the area-based weights.  The field-averaged bandpowers and covariance are then

\begin{equation}
D_b = \sum_{i}D_{b}^{i}w^{i}
\end{equation}
\begin{equation}
\textbf{C}_{bb^\prime} = \sum_{i}\textbf{C}_{bb^\prime}^{i}(w^{i})^2
\end{equation}
where 
\begin{equation}
w^{i} = \frac{A^i}{\sum_{i}A^i}
\end{equation}
is the area-based weight, and $A^i$ is the sum of the window of the $i^{\rm{th}}$ field.  We introduce the beam covariances by first calculating the covariance for each year, then adding in the beam covariance for that year, and finally combining the covariances of the two years.  The last step is to add the covariance due to the SPT calibration uncertainty, $\textbf{C}^{cal}_{ij} = \epsilon^{2}D_{i}D_{j}$, where $\epsilon=0.031$ corresponds to the 3.1\% uncertainty in the SPT power calibration discussed in Section~\ref{sec:calibration}.

\begin{figure*}[h]\centering
\includegraphics[width=1.0\textwidth]{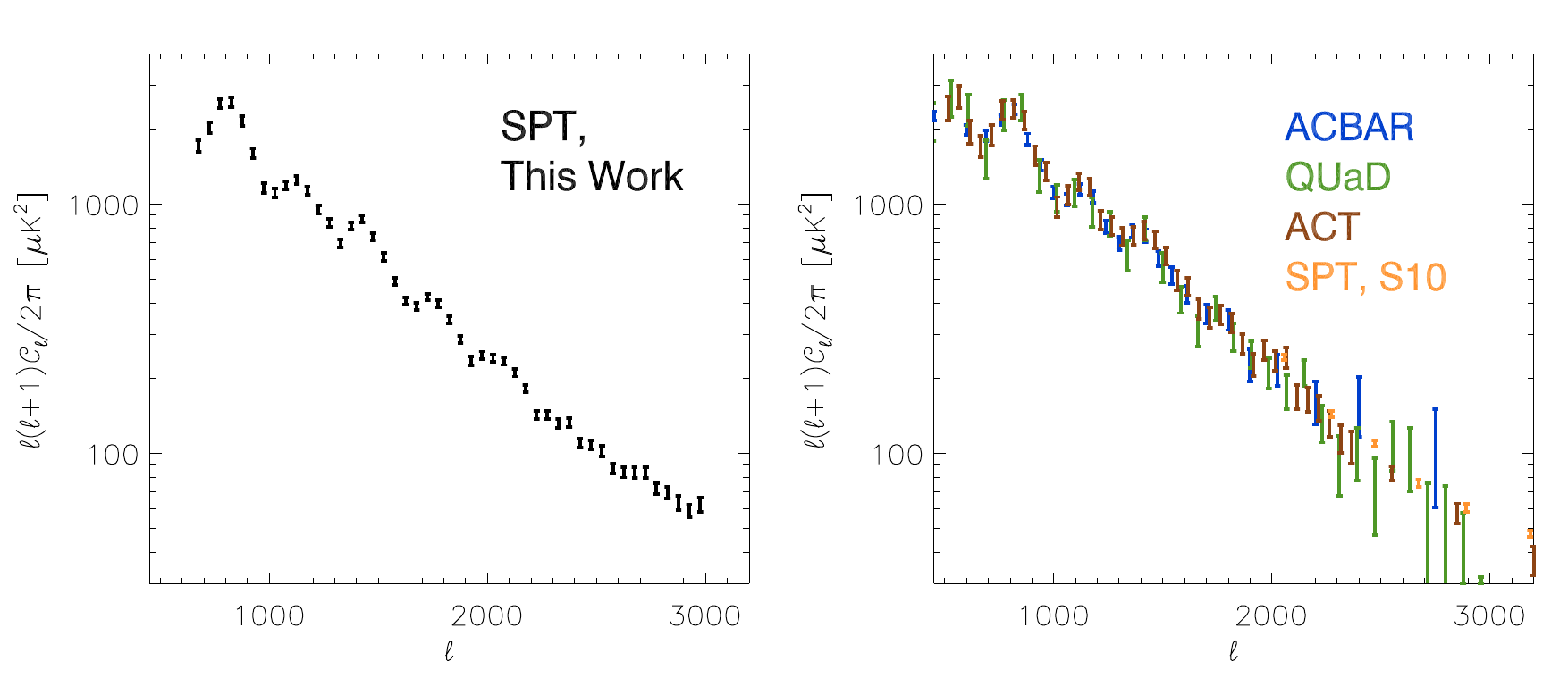}

  \caption[]{ The SPT power spectrum is shown in the left panel.  The peak at $\ell\sim800$ is the third acoustic peak.  For comparison we show in the right panel other recent measurements of the CMB damping tail from ACBAR \citep{reichardt09a}, QUaD \citep{friedman09, brown09}, ACT \citep{das10}, and SPT \citep{shirokoff11}.  The bandpower errors shown in these panels do not include beam or calibration uncertainties.  The ACT spectrum extends to $\ell=10,000$.  The previous SPT spectra, from \citet{lueker10} and \citet{shirokoff11}, spanned the angular range $2000 < \ell < 10,000$ and targeted secondary CMB anisotropy.}
    \label{fig:dl_all}
\end{figure*}

The final bandpowers are listed in Table~\ref{tab:dls} and shown in Figure~\ref{fig:dl_all}.

\subsection{Bandpower Window Functions \label{sec:bpwf}}

In order to allow for a theoretical power spectrum $C_\ell^{\rm th}$ to be compared to the SPT bandpowers $C_b$, we calculate the bandpower window functions ${\mathcal W}^b_\ell / \ell$, defined as

\begin{equation}
C_b^{\rm th} = ({\mathcal W}^b_\ell / \ell) C_\ell^{\rm th}.
\end{equation}

Following the formalism described in Section~\ref{sec:unbiased}, we can write this as

\begin{equation}
C_b^{\rm th} = (K^{-1})_{bb'} P_{b' \ell'} M_{\ell' \ell} F_\ell B_\ell^2 C_\ell^{\rm th},
\end{equation}
which implies that
\begin{equation}
 {\mathcal W}^b_\ell / \ell = (K^{-1})_{bb'} P_{b' \ell'} M_{\ell' \ell} F_\ell B_\ell^2.
\end{equation}

We calculate the bandpower window functions to be used with the final spectrum as the weighted average of the bandpower window functions from each field.

\begin{table*}
\begin{small}
\begin{center}
\begin{threeparttable}
\caption[SPT Bandpowers and Bandpower Errors]{SPT Bandpowers and Bandpower Errors}
\begin{tabular}{ | c c c | c c c | c c c | }
\hline
$\ell_{\rm center}$ & $D_\ell$ & $\sigma(D_\ell)$  & $\ell_{\rm center}$ & $D_\ell$  & $\sigma(D_\ell)$  & $\ell_{\rm center}$ & $D_\ell$  & $\sigma(D_\ell)$  \\
\hline
675 & 1710 & 95 & 1475 & 739 & 25 & 2275 & 143 & 5.4 \\
725 & 2010 & 98 & 1525 & 612 & 21 & 2325 & 132 & 5.1 \\
775 & 2530 & 110 & 1575 & 489 & 17 & 2375 & 133 & 5.0 \\
825 & 2560 & 110 & 1625 & 407 & 14 & 2425 & 110 & 4.5 \\
875 & 2150 & 93 & 1675 & 388 & 14 & 2475 & 108 & 4.4 \\
925 & 1600 & 69 & 1725 & 424 & 13 & 2525 & 102 & 4.7 \\
975 & 1160 & 51 & 1775 & 396 & 13 & 2575 & 86.6 & 3.8 \\
1025 & 1100 & 43 & 1825 & 343 & 11 & 2625 & 83.6 & 3.8 \\
1075 & 1190 & 46 & 1875 & 286 & 9.8 & 2675 & 83.4 & 4.1 \\
1125 & 1250 & 47 & 1925 & 234 & 8.8 & 2725 & 83.8 & 4.0 \\
1175 & 1130 & 43 & 1975 & 247 & 8.2 & 2775 & 71.9 & 3.6 \\
1225 & 946 & 35 & 2025 & 241 & 8.2 & 2825 & 69.5 & 3.9 \\
1275 & 839 & 29 & 2075 & 233 & 7.7 & 2875 & 63.1 & 4.1 \\
1325 & 696 & 27 & 2125 & 210 & 7.0 & 2925 & 58.9 & 3.5 \\
1375 & 813 & 27 & 2175 & 182 & 6.2 & 2975 & 62.0 & 4.1 \\
1425 & 869 & 28 & 2225 & 142 & 5.4 &  &  &  \\
\hline
\end{tabular}
\label{tab:dls}
\begin{tablenotes}
\item The SPT bandpowers and associated errors in units of $\microKsq$.  The errors do not include uncertainty in the SPT beam or calibration.
\end{tablenotes}
\end{threeparttable}
\end{center}
\end{small}
\end{table*}


\section{Cosmological Constraints}
\label{sec:cosmo}

The SPT power spectrum\footnote{Several of the data products presented in this work will be made available at http://pole.uchicago.edu/public/data/keisler11 and at http://lambda.gsfc.nasa.gov/product/spt .} described in the previous section should be dominated by primary CMB anisotropy and can be used to refine estimates of cosmological model parameters.  
In this section we constrain cosmological parameters using the SPT power spectrum in conjunction with data from the seven-year WMAP data release (WMAP7, \citealt{larson10})\footnote{We note that there is a small covariance between the SPT and WMAP bandpowers due to common sky coverage, but that it is negligibly small.  The composite error has been underestimated by $< 1\%$ across the overlapping $\ell$ range.} and, in some cases, in conjunction with low-redshift measurements of the Hubble constant $H_0$ using the Hubble Space Telescope \citep{riess11} and the baryon acoustic oscillation (BAO) feature using SDSS and 2dFGRS data \citep{percival10}.  In the analyses that follow, the label ``$H_0$+BAO'' implies that the following Gaussian priors have been applied: $H_0 = 73.8 \pm 2.4$ km s$^{-1}$ Mpc$^{-1}$; $r_s/D_{V}(z=0.2)=0.1905 \pm 0.0061$; and $r_s/D_V(z	=0.35)=0.1097 \pm 0.0036$; where $r_s$ is the comoving sound horizon size at the baryon drag epoch, $D_V(z) \equiv [(1 + z)^2 D^2_A(z)cz/H(z)]^{1/3}$, $D_A(z)$ is the angular diameter distance, and $H(z)$ is the Hubble parameter.  The inverse covariance matrix given in Eq. 5 of \citet{percival10} is used for the BAO measurements.

\subsection{Cosmological Model}
\label{sec:cosmo_model}

We fit the bandpowers to a model that includes four components: 

\begin{itemize}

\item Primary CMB.  We use the standard, six-parameter, spatially flat, $\Lambda$CDM cosmological model to predict the power from primary CMB anisotropy. The six parameters are the baryon density $\Omega_b h^2$, the density of cold dark matter $\Omega_c h^2$, the optical depth of reionization $\tau$, the angular scale of the sound horizon at last scattering $\theta_s$, the amplitude of the primordial scalar fluctuations (at pivot scale $k_0=0.002$ Mpc$^{-1}$) $\Delta^2_{R}$, and the spectral index of the scalar fluctuations $n_s$.  The effects of gravitational lensing on the power spectrum of the CMB are calculated using a cosmology-dependent lensing potential \citep{lewis06}.

\item ``Poisson'' point source power.  Our model includes a term to account for the shot-noise fluctuation power from randomly distributed, emissive galaxies.  This term is constant in $C_\ell$ and goes as $D^{\rm PS}_\ell \propto \ell^2$.

\item ``Clustered'' point source power.  Our model includes a term to account for the clustering of emissive galaxies.  For this clustering contribution we use the template $D_\ell^{\rm CL} \propto f_\ell$, where $f_\ell=1$ for $\ell<1500$, and $f_\ell=(\frac{\ell}{1500})^{0.8}$ for $\ell \ge 1500$.  This shape is designed to approximate the shape of the clustering power in both the linear and non-linear regimes \citep{shirokoff11, millea11}.  We find that our cosmological constraints are not sensitive to the details of this shape.  For example, there are no significant changes in the cosmological results if this shape is changed to a pure power law $D_\ell^{\rm CL} \propto \ell^{0.8}$.

\item SZ power.  Our model includes a term to account for power from the thermal and kinetic SZ effects.  At the angular scales considered here, the two effects are expected to have similar shapes in $\ell$-space.  We therefore adopt the thermal SZ template provided in \cite{sehgal10}, which has a shape similar to the templates predicted by more recent models \citep{trac11, shaw10, battaglia10}, to account for the total SZ power.

\end{itemize}

For the purposes of this analysis, the primary CMB encodes the cosmological information, while the last three components, the ``foreground'' terms, are nuisance parameters.  The foreground terms are used only when calculating the SPT likelihood; they are not used when calculating the WMAP likelihood.  In our baseline model, we apply a Gaussian prior on the amplitude of each of the foreground terms.  The prior on the Poisson power is $D_{3000}^{\rm PS} = 19.3 \pm 3.5~\microKsq$ and is based on the power from sources with $\rm{S}_{\rm 150 GHz} < 6.4~\rm{mJy}$, as measured in \cite{shirokoff11}, and the power from sources with $6.4~\rm{mJy} < S_{\rm 150 GHz} < 50~\rm{mJy}$, as measured in \cite{vieira10} and \cite{marriage11}.  The priors on the clustered power and SZ power are $D_{3000}^{\rm CL} = 5.0 \pm 2.5~\microKsq$ and $D_{3000}^{\rm SZ} = 5.5 \pm 3.0~\microKsq$, as measured in \cite{shirokoff11}.  The widths of these priors span the modeling uncertainties in the relevant papers.  Finally, we require the foreground terms to be positive.  We find that our constraints on cosmological parameters do not depend strongly on these priors, as discussed below.

One foreground that we have not explicitly accounted for is the emission from cirrus-like dust clouds in our Galaxy.  Using a procedure similar to that described in \citet{hall10}, we cross-correlate the SPT maps with predictions for the galactic dust emission at $150$~GHz in the SPT fields using model 8 of \citet{finkbeiner99}.  We detect the galactic dust in the cross-correlation and estimate the power from galactic dust in the field-averaged SPT spectrum to be $D_\ell = (1.4 \pm 0.4) \left(\frac{\ell}{1000}\right)^{-1.2} \microKsq$.  This is small compared to the SPT bandpower errors; subtracting this component changes the $\chi^2$ by 0.04 when all other parameters are kept fixed and by a negligible amount if the other foreground parameters are allowed to move by amounts that are small compared to the widths of their priors.  We conclude that galactic dust does not significantly contaminate the SPT power spectrum.

The total number of parameters in our baseline model is nine: six for the primary CMB and three for foregrounds.  The nine-dimensional space is explored using a Markov Chain Monte Carlo (MCMC) technique.  We use the CosmoMC\footnote{http://cosmologist.info/cosmomc/} software package \citep{lewis02b}, which itself uses the CAMB\footnote{http://camb.info/ .  We use RECFAST Version 1.5.} software package \citep{lewis99} to calculate the lensed CMB power spectra.  The CMB temperature and polarization spectra are calculated by CAMB for each cosmology.  These spectra are passed to the likelihood software provided by the WMAP team\footnote{We use Version 4.1 of the WMAP likelihood code, available at http://lambda.gsfc.nasa.gov/ .  In order to be consistent with our SPT+WMAP Markov chains, we recalculate all WMAP-only Markov chains rather than use those available at http://lambda.gsfc.nasa.gov/ .} to calculate the WMAP likelihood.  The SPT likelihood is calculated using the bandpowers, covariances, and bandpower window functions described in Section~\ref{sec:powspec}.  The age of the universe is required to be between 10 and 20 Gyr, and the Hubble constant is required to be $0.4 < h < 1.0$ where $H_0=h~100~\rm{km~s^{-1}~Mpc^{-1}}$.  We assume that neutrinos are massless.

Before combining the SPT and WMAP likelihoods, we first check that they independently give consistent constraints on the six cosmological parameters.  Because the scalar amplitude $\Delta^2_{R}$ and the optical depth $\tau$ are completely degenerate given only the high-$\ell$ SPT bandpowers, we fix $\tau=0.088$ for the cosmological model that is constrained using only SPT data.  We find that WMAP alone gives $\{100\Omega_bh^2, \Omega_ch^2, 100\theta_s, n_s, 10^9 \Delta^2_{R}\}$ = $\{2.24\pm0.056, 0.112\pm0.0054, 1.039\pm0.0027, 0.971\pm0.014, 2.42\pm0.11\}$, while SPT alone gives $\{2.19\pm0.18, 0.110\pm0.013, 1.043\pm0.0022, 0.953\pm0.048, 2.49\pm0.49\}$.  Thus the two likelihoods are consistent with each other, and we proceed to combine them.

\begin{figure*}[p]\centering
\includegraphics[width=0.8\textwidth]{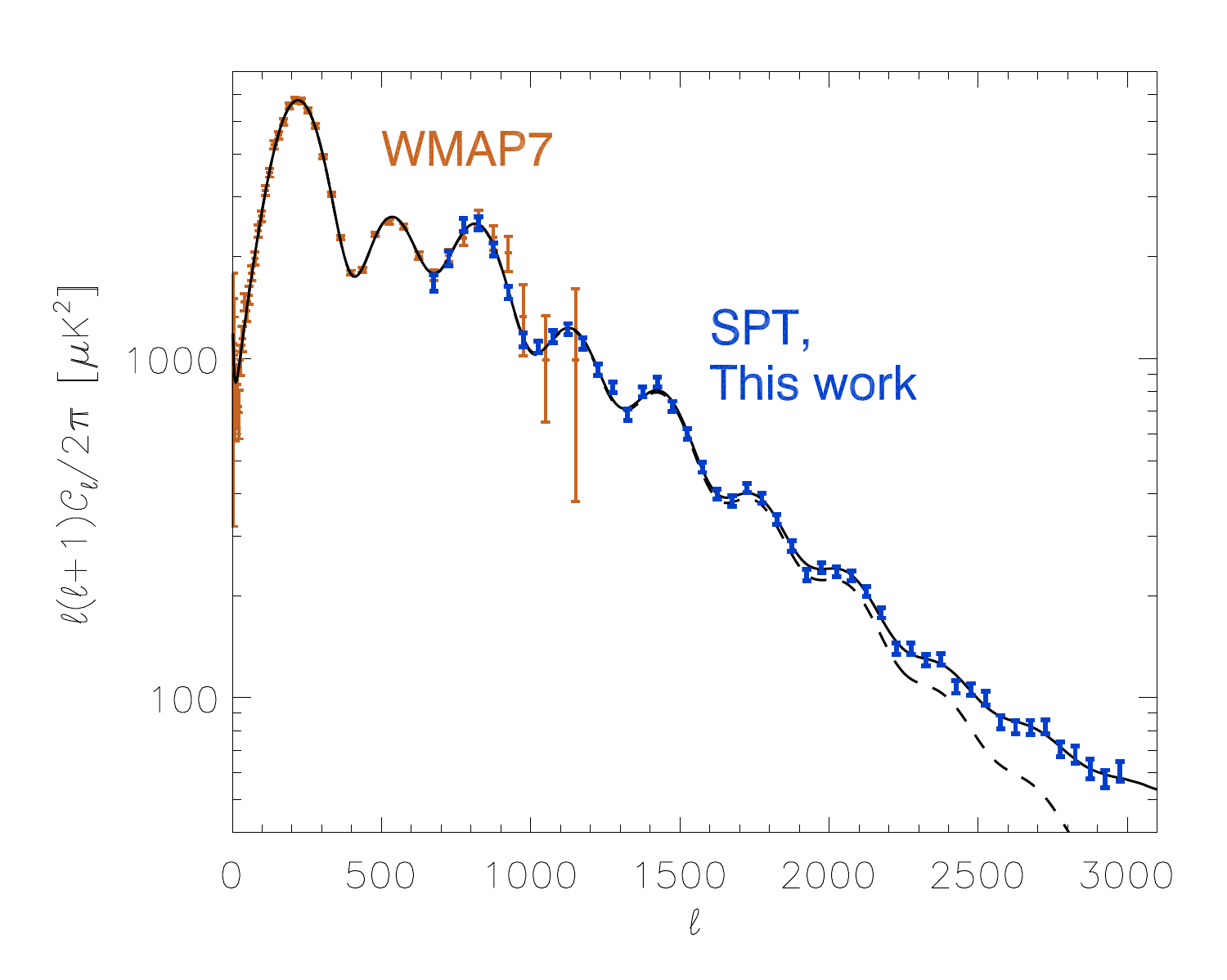}
  \caption[]{ The SPT bandpowers, WMAP bandpowers, and best-fit $\Lambda$CDM theory spectrum shown with dashed (CMB) and solid (CMB+foregrounds) lines.  The bandpower errors do not include beam or calibration uncertainties.}
  \label{fig:bestfit}
\end{figure*}

The best-fit model from the joint SPT+WMAP likelihood is shown in Figure~\ref{fig:bestfit}.  The baseline model provides a good fit to the SPT data.  The $\chi^2/\rm{dof}$ is $35.5/38$ (PTE=0.58) if the six cosmological parameters are considered free and $35.5/44$ (PTE=0.82) if the six cosmological parameters are considered to be essentially fixed by the WMAP data.

\begin{figure*}[p]\centering
\includegraphics[width=1.0\textwidth]{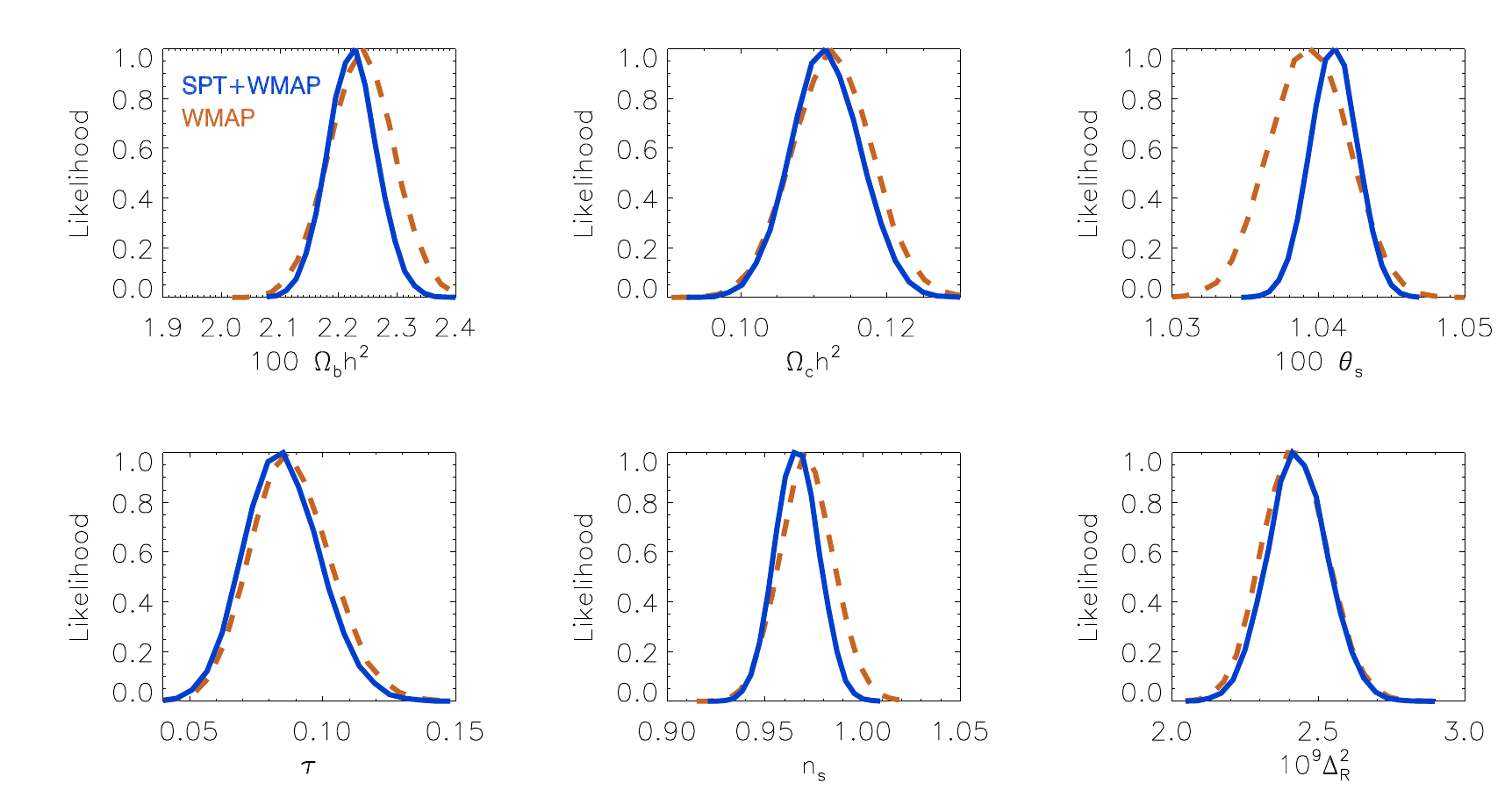}
  \caption[]{ The one-dimensional marginalized constraints on the six cosmological parameters in the baseline model.  The constraints from SPT+WMAP are shown by the blue solid lines, while the constraints from WMAP alone are shown by the orange dashed lines.}
  \label{fig:baseline}
\end{figure*}

The marginalized likelihood distributions for the six cosmological parameters are shown in Figure~\ref{fig:baseline}.  The addition of the SPT data improves the constraints on $\Omega_b h^2$ and $n_s$ by $\sim$25\% and the constraint on $\theta_s$ by nearly a factor of two.  The parameter constraints for the baseline model are summarized in Table~\ref{tab:params}, and constraints on this model using SPT+WMAP+$H_0$+BAO are given in Table~\ref{tab:params_hbao}.

The scalar spectral index $n_s$ is less than one in simple models of inflation \citep{linde08}.  Recent measurements of $n_s$ come from WMAP7, $n_s=0.967\pm0.014$ \citep{komatsu11}, ACBAR+QUaD+WMAP7, $n_s = 0.966^{+0.014}_{-0.013}$ \citep{komatsu11}, and ACT+WMAP7, $n_s=0.962\pm0.013$ \citep{dunkley10}.  The SPT+WMAP constraint is

\begin{equation}
n_s=0.9663\pm0.0112 \,.
\end{equation}

This is a 3.0$\sigma$ preference for $n_s<1$ over the Harrison-Zel'dovich-Peebles index, \ns$=1$.  This constraint is not significantly altered if we double the width of the priors on the foreground terms, in which case \ns$=0.9666\pm0.0112$.  The constraint is also robust to doubling our uncertainties on the SPT beam functions or SPT calibration, in which cases \ns$=0.9671\pm0.0113$ and \ns$=0.9661\pm0.0111$, respectively.  When the $H_0$ and BAO data are included, the constraint is \ns$=0.9668\pm0.0093$, a 3.6$\sigma$ preference for $n_s<1$.

\begin{table*}[p]
\begin{center}
\begin{threeparttable}
\caption[Constraints on Cosmological Parameters using SPT+WMAP]{Constraints on Cosmological Parameters using SPT+WMAP}
\begin{tabular}{ | l  c  c  c  c  c  c c | }
\hline \hline
 & & $\Lambda$CDM & $\Lambda$CDM & $\Lambda$CDM & $\Lambda$CDM & $\Lambda$CDM & $\Lambda$CDM \\
 & & & +~\alens & $+r$ & +~\nrun & +~\yhe & +~\neff \\
  \hline \hline
Primary & $100\Omega_{b}h^2$ & $2.22\pm0.042$ & $2.22\pm0.044$ & $2.24\pm0.045$ & $2.21\pm0.043$ & $2.26\pm0.048$ & $2.27\pm0.054$\\
Parameters & $\Omega_{c}h^2$ & $0.112\pm0.0048$ & $0.112\pm0.0050$ & $0.109\pm0.0050$ & $0.117\pm0.0059$ & $0.114\pm0.0051$ & $0.125\pm0.012$\\
 & $100\theta_s$ & $1.041\pm0.0016$ & $1.041\pm0.0016$ & $1.041\pm0.0016$ & $1.041\pm0.0017$ & $1.043\pm0.0020$ & $1.040\pm0.0019$\\
 & \ns & $0.9663\pm0.0112$ & $0.9655\pm0.0114$ & $0.9743\pm0.0128$ & $0.9732\pm0.0120$ & $0.9793\pm0.0140$ & $0.9874\pm0.0193$\\
 & $\tau$ & $0.0851\pm0.014$ & $0.0853\pm0.014$ & $0.0860\pm0.014$ & $0.0909\pm0.016$ & $0.0884\pm0.015$ & $0.0883\pm0.015$\\
 & $10^9 \Delta^2_{R}$ & $2.43\pm0.10$ & $2.44\pm0.11$ & $2.36\pm0.11$ & $2.37\pm0.11$ & $2.39\pm0.10$ & $2.37\pm0.11$\\
\hline
Extension & $A^{0.65}_{L}$ & --- & $0.94\pm0.15$ & --- & --- & --- & ---\\
Parameters & $r$ & --- & --- & $<0.21$ & --- & --- & ---\\
 & \nrun & --- & --- & --- & $-0.024\pm0.013$ & --- & ---\\
 & \yhe & ($0.2478\pm0.0002$) & ($0.2478\pm0.0002$) & ($0.2479\pm0.0002$) & ($0.2477\pm0.0002$) & $0.296\pm0.030$ & ($0.2579\pm0.008$)\\
 & \neff & (3.046) & (3.046) & (3.046) & (3.046) & (3.046) & $3.85\pm0.62$\\
\hline
Derived & $\sigma_8$ & ($0.814\pm0.024$) & ($0.816\pm0.025$) & ($0.805\pm0.025$) & ($0.832\pm0.027$) & ($0.837\pm0.029$) & ($0.859\pm0.043$)\\
\hline
 & $\chi^2_{\rm min}$ & 7506.5 & 7506.2 & 7506.4 & 7503.6 & 7504.4 & 7505.5\\
  \hline \hline
\end{tabular}
\label{tab:params}
\begin{tablenotes}
\item The constraints on cosmological parameters using SPT+WMAP7.  We report the mean of the likelihood distribution and the symmetric 68\% confidence interval about the mean.  We report the 95\% upper limit on the tensor-to-scalar ratio $r$.  Parameters labeled with ``---'' were held at their default values: $A_L=1, r=0,$ and $\nrun=0$.
\end{tablenotes}
\end{threeparttable}
\end{center}
\end{table*}

\begin{table*}[p]
\begin{center}
\begin{threeparttable}
\caption[Constraints on Cosmological Parameters using SPT+WMAP+$H_0$+BAO]{Constraints on Cosmological Parameters using SPT+WMAP+$H_0$+BAO}
\begin{tabular}{ | l  c  c  c  c  c  c c | }
\hline \hline
 & & $\Lambda$CDM & $\Lambda$CDM & $\Lambda$CDM & $\Lambda$CDM & $\Lambda$CDM & $\Lambda$CDM \\
 & & & +~\alens & $+r$ & +~\nrun & +~\yhe & +~\neff \\
  \hline \hline
Primary & $100\Omega_{b}h^2$ & $2.23\pm0.038$ & $2.22\pm0.039$ & $2.24\pm0.040$ & $2.23\pm0.040$ & $2.27\pm0.044$ & $2.26\pm0.042$\\
Parameters & $\Omega_{c}h^2$ & $0.112\pm0.0028$ & $0.112\pm0.0029$ & $0.112\pm0.0030$ & $0.114\pm0.0031$ & $0.114\pm0.0032$ & $0.129\pm0.0093$\\
 & $100\theta_s$ & $1.041\pm0.0015$ & $1.041\pm0.0016$ & $1.041\pm0.0015$ & $1.041\pm0.0016$ & $1.043\pm0.0020$ & $1.039\pm0.0017$\\
 & \ns & $0.9668\pm0.0093$ & $0.9659\pm0.0095$ & $0.9711\pm0.0099$ & $0.9758\pm0.0111$ & $0.9814\pm0.0126$ & $0.9836\pm0.0124$\\
 & $\tau$ & $0.0851\pm0.014$ & $0.0852\pm0.014$ & $0.0842\pm0.014$ & $0.0934\pm0.016$ & $0.0890\pm0.015$ & $0.0859\pm0.014$\\
 & $10^9 \Delta^2_{R}$ & $2.43\pm0.082$ & $2.44\pm0.085$ & $2.39\pm0.088$ & $2.35\pm0.095$ & $2.39\pm0.085$ & $2.41\pm0.084$\\
\hline
Extension & $A^{0.65}_{L}$ & --- & $0.95\pm0.15$ & --- & --- & --- & ---\\
Parameters & $r$ & --- & --- & $<0.17$ & --- & --- & ---\\
 & \nrun & --- & --- & --- & $-0.020\pm0.012$ & --- & ---\\
 & \yhe & ($0.2478\pm0.0002$) & ($0.2478\pm0.0002$) & ($0.2478\pm0.0002$) & ($0.2478\pm0.0002$) & $0.300\pm0.030$ & ($0.2581\pm0.005$)\\
 & \neff & (3.046) & (3.046) & (3.046) & (3.046) & (3.046) & $3.86\pm0.42$\\
\hline
Derived & $\sigma_8$ & ($0.818\pm0.019$) & ($0.818\pm0.019$) & ($0.816\pm0.019$) & ($0.824\pm0.020$) & ($0.841\pm0.024$) & ($0.871\pm0.033$)\\
\hline
 & $\chi^2_{\rm min}$ & 7510.7 & 7510.6 & 7510.7 & 7507.8 & 7508.0 & 7507.4\\
  \hline \hline
\end{tabular}
\label{tab:params_hbao}
\begin{tablenotes}
\item The constraints on cosmological parameters using SPT+WMAP7+$H_0$+BAO.  We report the mean of the likelihood distribution and the symmetric 68\% confidence interval about the mean.  We report the 95\% upper limit on the tensor-to-scalar ratio $r$.  Parameters labeled with ``---'' were held at their default values: $A_L=1, r=0,$ and $\nrun=0$.
\end{tablenotes}
\end{threeparttable}
\end{center}
\end{table*}

\begin{table*}
\begin{center}
\begin{threeparttable}
\caption[Constraints on Model Extensions using Recent CMB Datasets]{Constraints on Model Extensions using Recent CMB  Datasets}
\begin{tabular}{ | l c  c  c  c | }
\hline 
 & WMAP7 & ACBAR+QUaD+WMAP7 & ACT+WMAP7 & SPT+WMAP7 \\
\hline \hline
$r$ & $<0.7$ & $<0.33$ & $<0.25$ & $<0.21$ \\
\nrun & [-0.084, 0.020] & [-0.084, 0.003] & $-0.034 \pm 0.018$ & $-0.024 \pm 0.013$ \\
\yhe & $< 0.51$ & $0.326 \pm 0.075$ & $0.313 \pm 0.044$ & $0.296 \pm 0.030$ \\
\neff & $> 2.7$ & --- & $5.3 \pm 1.3$ & $3.85 \pm 0.62$ \\
\hline
\end{tabular}
\label{tab:recent_cmb}
\begin{tablenotes}
\item The constraints on cosmological parameters in certain model extensions using recent CMB datasets.  We use WMAP7 \citep{larson10, komatsu11}, ACBAR \citep{reichardt09a}, QUaD \citep{brown09}, ACT \citep{das10}, and SPT (this work).  All upper and lower limits and 
all two-sided limits (shown in brackets) are 95\%.
\end{tablenotes}
\end{threeparttable}
\end{center}
\end{table*}


\subsection{Model Extensions}
\label{sec:extensions}
In this section we consider extensions to our baseline model.  These models continue to use a spatially flat, $\Lambda$CDM cosmological model, but allow a previously fixed parameter---the strength of gravitational lensing, the amplitude of tensor fluctuations, the running of the spectral index, the primordial helium abundance, or the number of relativistic species---to vary freely.  The structure of this section closely follows the clear presentation and discussion of the ACT+WMAP constraints on parameter extensions by \citet{dunkley10}, and therefore allows a straightforward comparison of these similar datasets.  We summarize constraints on these extension parameters using recent CMB datasets in Table~\ref{tab:recent_cmb}.

We note that we have also considered extensions with a free dark energy equation of state $w$ or with massive neutrinos, and found that the addition of SPT data did not significantly improve upon the constraints on these models from WMAP alone.

\subsubsection{Gravitational Lensing}
\label{sec:lens}

The paths of CMB photons are distorted by the gravity of intervening matter as they travel from the surface of last scattering to us, a process referred to as gravitational lensing.  The typical deflection angle is a few arcminutes, and the deflections are coherent over degree scales.  Lensing encodes information on the distribution of matter at intermediate redshifts, and this information can be partially recovered using correlations of the CMB temperature and polarization fields \citep{bernardeau97, zaldarriaga99, hu01b, hu02a}.

Previous efforts to detect lensing of the CMB include $\sim$$3\sigma$ detections from correlating quadratic reconstructions of the lensing field from the CMB with other mass tracers \citep{smith07, hirata08},  $\lesssim3\sigma$ detections from the temperature power spectrum \citep{reichardt09a, calabrese08, das10}, a $2\sigma$ detection from the four-point function of the temperature field \citep{smidt11}, and a $4\sigma$ detection from the four-point function of the temperature field \citep{das11}.

The CMB temperature power spectrum is altered by lensing at the few percent level.  The acoustic peak structure is smoothed, and power is preferentially added to smaller angular scales.  The SPT bandpowers accurately measure the acoustic peaks in the damping tail and should be sensitive to these effects.  The effect of lensing on the CMB temperature power spectrum is largely captured by considering large-scale structure in the linear regime \citep{lewis06}.  The corrections due to non-linear structures are small compared to the SPT bandpower errors and are not considered in this analysis.

As a simple measure of the preference for lensing, we fit the SPT+WMAP bandpowers to a model in which the CMB is not lensed but that is otherwise identical to our baseline model.  All parameters are free in this no-lensing model.  We compare the likelihood of the best-fit no-lensing model to the likelihood of the best-fit lensed model.  We find $\Delta\chi^2 = 2(\ln \mathcal{L}^{\rm lens} - \ln \mathcal{L}^{\rm no~lens}) = 23.6$, corresponding to a $4.9\sigma$ preference for the lensed model.  This preference remains if we double the widths of the priors on the foreground parameters, in which case $\Delta\chi^2=23.0$, or if we double the SPT beam uncertainty, in which case $\Delta\chi^2=22.9$.  We note that the distortion to the power spectrum caused by adopting the flat-sky approximation on the SPT maps is small compared to the effect of gravitational lensing on the power spectrum.  We estimate that correcting for this distortion would change the SPT likelihood by $\Delta\chi^2 <0.04$.

To better quantify the strength of lensing preferred by the data, we consider the parameter \alens\, which rescales the lensing potential power spectrum,

\begin{equation}
C_\ell^{\phi\phi} \rightarrow A_{L} C_\ell^{\phi\phi}.
\end{equation} 
As in all of our models, $C_\ell^{\phi\phi}$ is calculated in a cosmology-dependent manner at each point in the Markov chain.  The \alens\ parameter is phenomenological and we allow it to be negative.  In the standard scenario, $\alens=1$, while $\alens=0$ corresponds to no lensing.

This parameter has been constrained using measurements of the CMB damping tail in conjunction with WMAP data, and more recently by measuring the lensing signal encoded in the CMB temperature four-point function.  \citet{reichardt09a} used ACBAR data in conjunction with five-year WMAP data \citep{nolta09} and found $\alens=1.60^{+0.55}_{-0.26}$, while \citet{calabrese08} found $\alens=3.0 \pm 0.9$ using the same datasets.  \citet{das10} used power spectra from ACT+WMAP7 to measure $\alens=1.3 \pm 0.5$, and \citet{das11} used the four-point function of the ACT temperature maps to measure $\alens=1.16 \pm 0.29$.

\begin{figure*}[h]\centering
\includegraphics[]{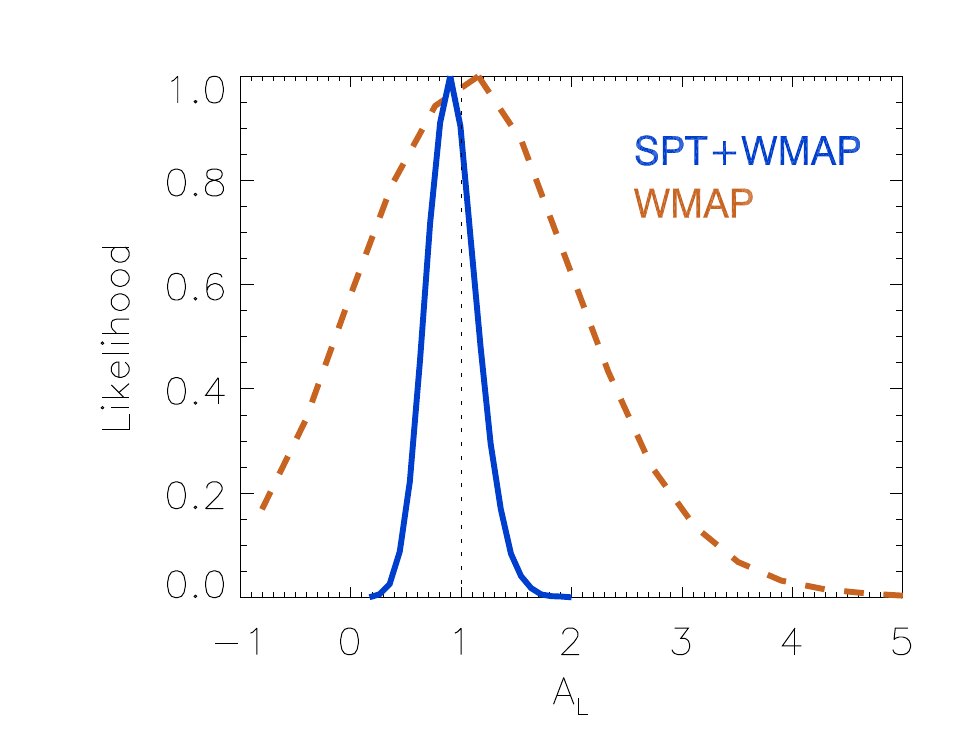}
  \caption[]{ The one-dimensional marginalized constraint on the gravitational lensing parameter \alens.  This parameter rescales the lensing potential power spectrum as $C_\ell^{\phi\phi} \rightarrow A_{L} C_\ell^{\phi\phi}$.}
  \label{fig:alens}
\end{figure*}

The constraints on \alens\ from SPT+WMAP7 are shown in Figure~\ref{fig:alens}.  While the constraint on \alens\ is non-Gaussian in shape, we find that the constraint on $A^{0.65}_{L}$ (which still has an expectation value of 1) is approximately Gaussian.  With SPT+WMAP7, we find\footnote{Note that this constraint, along with the other constraints on $A^{0.65}_{L}$ listed in this work, implicitly assumes a uniform prior on $A_{L}$.  However, the result does not change significantly if we modify the prior to be uniform in $A^{0.65}_{L}$ instead, in which case $A^{0.65}_{L} = 0.93 \pm 0.15$.}

\begin{equation}
A^{0.65}_{L} = 0.94 \pm 0.15.
\end{equation}
This constraint does not change significantly if we double the width of the priors on the foreground terms, in which case $A^{0.65}_{L} = 0.94 \pm 0.15$.  The SPT+WMAP7 data reject a non-lensed CMB and are consistent with the expected level of lensing.  This provides a consistency check on the standard picture of large-scale structure formation.  Ongoing and future measurements of CMB lensing will move beyond this consistency check and constrain parameters that affect the growth of large-scale structure, such as the properties of dark energy and the sum of the neutrino masses \citep{smith06b, sherwin11}.

\subsubsection{Tensor Perturbations}
\label{sec:r}

Inflation is expected to produce primordial tensor perturbations (\ie\ gravitational waves).  These perturbations imprint potentially detectable effects onto the CMB temperature and polarization spectra.  The amplitude of the tensor spectrum is often given in terms of the the tensor-to-scalar ratio, $r=\Delta_h^2(k_0)/\Delta_R^2(k_0)$ with pivot scale $k_0=0.002$ Mpc$^{-1}$.\footnote{We assume that the spectral index of the tensor perturbations is $n_t=-r/8$.}  A detection of $r$ would provide an extremely interesting window onto the energy scale of inflation.

Measurements of the B-mode polarization at low multipoles will ultimately provide the strongest constraints on $r$.  To date, the best constraint on $r$ from B-mode polarization comes from the BICEP experiment \citep{chiang10}, giving $r<0.7$ (95\% CL).  Stronger constraints are currently placed using WMAP's measurement of the temperature and polarization spectra \citep{komatsu11}, which give $r<0.36$ (95\% CL).

\begin{figure*}[h]\centering
\includegraphics[width=1.0\textwidth]{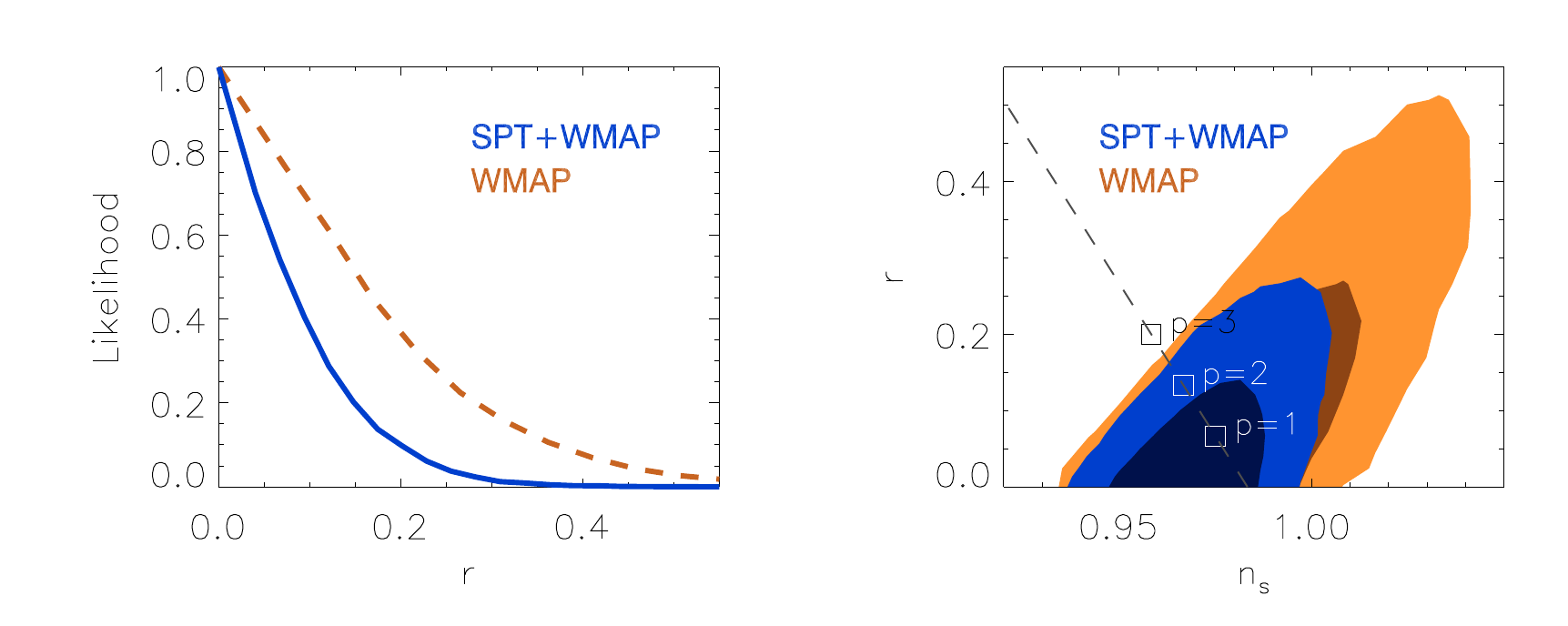}
  \caption[]{ The one-dimensional marginalized constraint on the tensor-to-scalar ratio $r$ (left) and the two-dimensional constraint on $r$ and the spectral index \ns\ (right).  The dashed line shows predictions for $r$ and \ns\ from chaotic inflationary models with inflaton potential $V(\phi) \propto \phi^p$ and $N=60$, where $N$ is the number of e-folds between the epoch when modes that are measured by SPT and WMAP exited the horizon during inflation and the end of inflation.  The two-dimensional contours show the 68\% and 95\% confidence intervals.}
  \label{fig:r}
\end{figure*}

The constraint on $r$ can be improved indirectly with the addition of small-scale CMB measurements.  The CMB power at low multipoles increases as $r$ increases, but this effect can be partially cancelled by increasing $n_s$ and decreasing $\Delta^2_{R}$.  The small-scale CMB measurements help to break these degeneracies, as demonstrated in \citet{komatsu11}, which found $r<0.33$ (95\% CL) using ACBAR+QUaD+WMAP7, and in \citet{dunkley10}, which found $r<0.25$ (95\% CL) using ACT+WMAP7.  The SPT+WMAP7 data constrain $r$ to be 

\begin{equation}
r < 0.21~(95\%\ \rm{CL}).
\end{equation}
When the $H_0$ and BAO data are added, the constraint improves to $r < 0.17~(95\%\ \rm{CL})$.  Figure~\ref{fig:r} shows the one-dimensional marginalized constraint on $r$ and the two-dimensional constraint for $r$ and the spectral index $n_s$.  We show the predictions for $r$ and $n_s$ from chaotic inflationary models \citep{linde83} with inflaton potential $V(\phi) \propto \phi^{p}$ and $N=60$, where $N$ is the number of e-folds between the epoch when modes that are measured by SPT and WMAP exited the horizon during inflation and the end of inflation.  These models predict $r=4p/N$ and $n_s=1-(p+2)/2N$.  Models with $p\ge3$ are disfavored at more than 95\% confidence for $N \le 60$.

\subsubsection{Running of the Spectral Index}
\label{sec:running}

The power spectrum of primordial scalar fluctuations is typically parametrized as a power law,

\begin{equation}
\Delta^{2}_{R}(k) = \Delta^{2}_{R}(k_0)\left(\frac{k}{k_0}\right)^{n_s-1}.
\end{equation}
In this section we allow the power spectrum to depart from a pure power law.  The ``running'' of the spectral index is parametrized as\footnote{The factor of $1/2$ is due to considering the effective change in $n_s$ in $d\ln\Delta^{2}_{R}/d\ln k$.}

\begin{equation}
\Delta^{2}_{R}(k) = \Delta^{2}_{R}(k_0)\left(\frac{k}{k_0}\right)^{n_s-1+\frac{1}{2}\ln(k/k_0) dn_s/d\ln k}.
\end{equation}

\begin{figure*}[h]\centering
\includegraphics[]{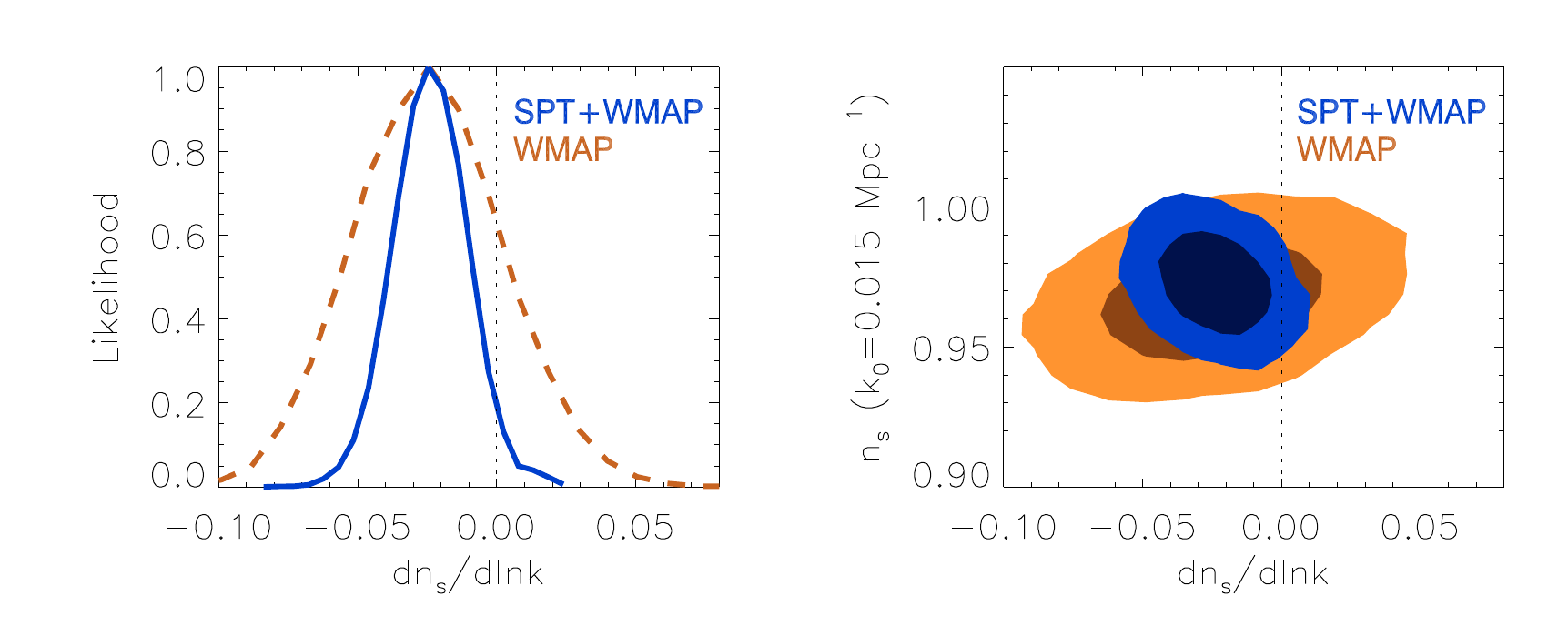}
  \caption[]{ The one-dimensional marginalized constraint on the running of the spectral index \nrun\ (left) and the two-dimensional constraint on \nrun\ and the spectral index \ns\ (right).  The two-dimensional contours show the 68\% and 95\% confidence intervals.}
  \label{fig:nrun}
\end{figure*}

The running parameter \nrun\ is predicted to be small by most inflationary theories, and a detection of a non-zero \nrun\ could provide information on the inflationary potential \citep{kosowsky95}.  Recent CMB constraints on the running include $-0.084<\nrun<0.020$ (95\% CL) from WMAP7 \citep{komatsu11}, $-0.084<\nrun<0.003$ (95\% CL) from ACBAR+QUaD+WMAP7 \citep{komatsu11}, and $\nrun=-0.034 \pm 0.018$ from ACT+WMAP7 \citep{dunkley10}.  The SPT+WMAP7 data constrain \nrun\ to be 

\begin{equation}
dn_s/d\ln k = -0.024 \pm 0.013.
\end{equation}
The data mildly prefer, at 1.8$\sigma$, a negative spectral running.  The constraint is $dn_s/d\ln k = -0.020 \pm 0.012$, a 1.7$\sigma$ preference for negative running, when the $H_0$ and BAO data are added.  As discussed in Section~\ref{sec:discuss_damping} and shown in Table~\ref{tab:params_hbao_clusters}, the constraint is $dn_s/d\ln k = -0.017 \pm 0.012$, a 1.4$\sigma$ preference for negative running, when information from local galaxy clusters is added.  Figure~\ref{fig:nrun} shows the one-dimensional marginalized constraint on \nrun\ and the two-dimensional constraint for \nrun\ and the spectral index $n_s$.\footnote{The estimates for \nrun~and $n_s$ are highly correlated for the typical pivot scale, $k_0=0.002$ Mpc$^{-1}$.  As in \citet{dunkley10}, we have calculated the spectral index at a new, less correlated pivot scale $k_0=0.015$ Mpc$^{-1}$, where $n_s(k_0=0.015$ Mpc$^{-1}$) = $n_s(k_0=0.002$ Mpc$^{-1}$$) + \ln(0.015/0.002)dn_s/d\ln k$ \citep{cortes07}.}

\subsubsection{Primordial Helium Abundance}
\label{sec:helium}

When the universe cools to $T~\sim$ 0.1 MeV, light nuclei begin to form, a process known as big bang nucleosynthesis (BBN) \citep{schramm98, steigman07}.  The primordial abundance (mass fraction) of $^4$He is denoted as \yhe\ and is a function of baryon density and the expansion rate during BBN \citep{simha08}:

\begin{equation}
\label{eq:yp}
\yhe = 0.2485 + 0.0016[(273.9\Omega_bh^2-6)+100(S-1)], 
\end{equation}
where
\begin{equation}
S^2 = 1 + (7/43)(\neff-3.046).
\end{equation}
The $S^2$ factor generically accounts for any non-standard expansion rate prior to and during BBN, here parametrized in terms of the effective number of relativistic particle species \neff.  We calculate \yhe\ in this ``BBN-consistent'' manner in all of our models, unless noted otherwise.\footnote{We note that \yhe\ is calculated within CAMB using the PArthENoPE BBN code \citep{pisanti08}, and that the \yhe\ calculated with this code differs from the function given in Eq.~\ref{eq:yp} by $<1\%$ in the range $\Omega_{b}h^2=[0.021,0.023]$ and $\neff=[2,6]$.  We pass $\Delta N_\nu=(\neff-3.046)$ to PArthENoPE, such that our definition of \neff\ refers to \neff\ in the epoch after electron-positron annihilation at $T~\sim 0.5$ MeV.}

For our baseline model, we find $\yhe=0.2478 \pm 0.0002$.  This very small uncertainty\footnote{The $\sim$0.1\% uncertainty quoted here is the statistical uncertainty and is slightly smaller than the 0.2\% theoretical uncertainty on \yhe\ in the PArthENoPE code used in CAMB \citep{pisanti08}.} on \yhe\ is due only to the uncertainty in $\Omega_bh^2$, as \neff\ is fixed to its standard value of 3.046.  This constraint relies on standard BBN theory being correct.  We can consider an independent measurement of \yhe: the value of \yhe\ preferred by the CMB due solely to the effect of helium on the CMB damping tail.  Helium combines earlier than hydrogen, and thus more helium (at fixed baryon density) leads to fewer free electrons during hydrogen recombination.  This, in turn, leads to larger diffusion lengths for photons and less power in the CMB damping tail.

A simple test of the preference of the CMB data for non-zero primordial helium follows.  We compare the maximum likelihood in a model with no $^4$He to the maximum likelihood in our baseline model.  Using SPT+WMAP7, we find that the standard, BBN-consistent helium abundance is preferred over no helium at 7.7$\sigma$ ($\Delta\chi^2=58.8$).

\begin{figure*}[h]\centering
\includegraphics[]{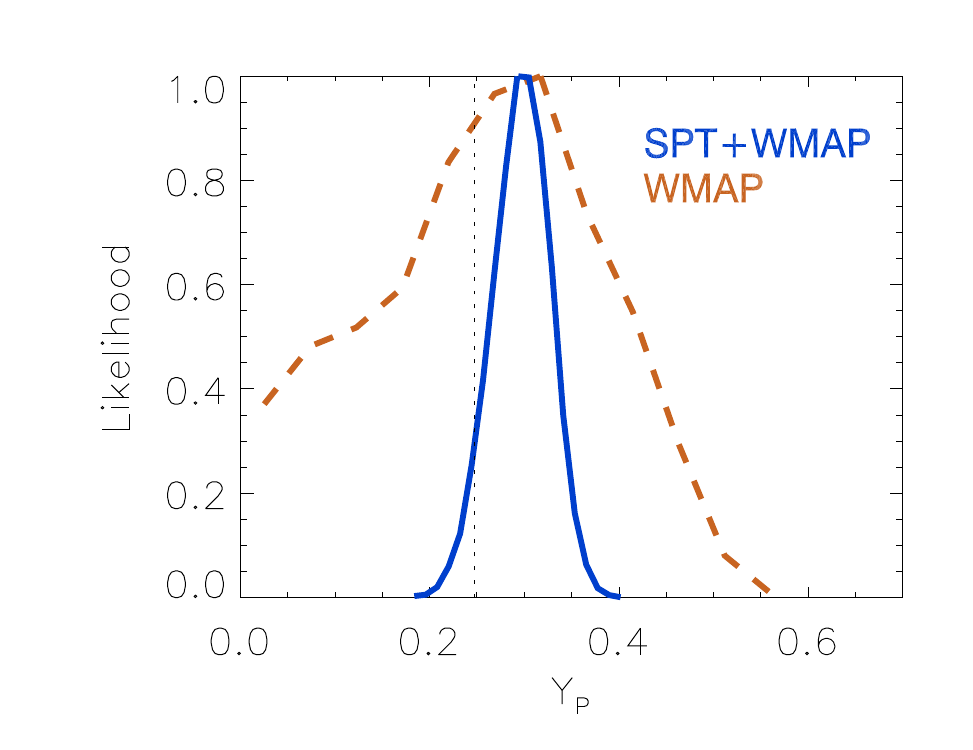}
  \caption[]{ The one-dimensional marginalized constraint on the primordial helium abundance \yhe.  The standard BBN value (\ie\ the value of \yhe\ in our best-fit baseline model, $\yhe=0.2478$) is shown by the dotted vertical line.}
  \label{fig:yhe}
\end{figure*}

We can extend this test by promoting \yhe\ to a free parameter.  In such a model, \yhe\ no longer obeys Eq.~\ref{eq:yp}, but rather is free to vary.  \citet{komatsu11} use WMAP7 to infer $\yhe<0.51$ (95\% CL) and ACBAR+QUaD+WMAP7 to measure $\yhe=0.326 \pm 0.075$.  \citet{dunkley10} use ACT+WMAP7 to measure $\yhe=0.313 \pm 0.044$.  The SPT+WMAP7 data constrain \yhe~to be 

\begin{equation}
\yhe = 0.296 \pm 0.030.
\end{equation}
The data mildly prefer, at 1.6$\sigma$, a value of \yhe\ that is larger than the value we obtain assuming standard BBN theory.  The constraint is $\yhe=0.300 \pm 0.030$, 1.7$\sigma$ higher than the standard BBN value, when the $H_0$ and BAO data are added.  As discussed in Section~\ref{sec:discuss_damping} and shown in Table~\ref{tab:params_hbao_clusters}, the constraint is $\yhe=0.288 \pm 0.029$, 1.4$\sigma$ higher than the standard BBN value, when information from local galaxy clusters is added.  Figure~\ref{fig:yhe} shows the one-dimensional marginalized constraint on \yhe.

The primordial $^4$He abundance may also be inferred from observations of low-metallicity extragalactic HII regions \citep{izotov07, peimbert07, izotov10, aver10, aver11}.  For example, \citet{aver11} provides an extensive analysis of the systematic uncertainties associated with these measurements and finds $\yhe=0.2609 \pm 0.0117$ (or $0.2573^{+0.0033}_{-0.0088}$ if the metallicity slope $d\yhe/dZ$ is required to be positive).  These values lie between, and are consistent with, the result from our baseline model, $\yhe=0.2478 \pm 0.0002$, and the result from our free-\yhe\ model, $\yhe=0.296 \pm 0.030$.

\subsubsection{Number of Relativistic Species}
\label{sec:neff}

In the standard theory of the early universe, there are three neutrino species that contribute $\sim$10\% of the energy density at recombination.  The effective number of particle species that are relativistic prior to and during recombination, \neff,\footnote{\neff\ is defined such that $\rho_\nu = \neff 7/8(4/11)^{4/3}\rho_\gamma$} is slightly higher (3.046) due to energy injection from electron-positron annihilation at the end of neutrino freeze-out \citep{dicus82, lopez99, mangano05}.  A significant detection of $\neff \ne 3.046$ could point to the presence of extra relativistic species in the early universe.

The addition of extra relativistic species increases the expansion rate during the radiation-dominated era.  If the parameters that are robustly measured by WMAP---$\Omega_bh^2$, $z_{\rm EQ}$,\footnote{$z_{\rm EQ}$ is the redshift at which matter and radiation have equal energy densities and satisfies (1+$z_{\rm EQ}$)=$\Omega_m/\Omega_r$, where $\Omega_m$ is the density of matter and $\Omega_r$ is the density of radiation (photons and neutrinos).} and $\theta_s$---are held fixed, then the main effect of this increased expansion rate is to increase the angular scale of photon diffusion and thereby lower power in the damping tail \citep{hu96a, bashinsky04, hou11}.  Conversely, measurements of the CMB damping tail can, in conjunction with WMAP, constrain the number of relativistic species.

A simple test of the preference of the CMB data for a non-zero number of relativistic species follows.  We compare the maximum likelihood in a model with \neff=0 to the maximum likelihood in our baseline model.  Using SPT+WMAP7, we find that the standard value of \neff=3.046 is preferred over no relativistic species at 7.5$\sigma$ ($\Delta\chi^2=56.3$).  The CMB data strongly prefer the existence of neutrinos over no neutrinos.

\begin{figure*}[h]\centering
\includegraphics[]{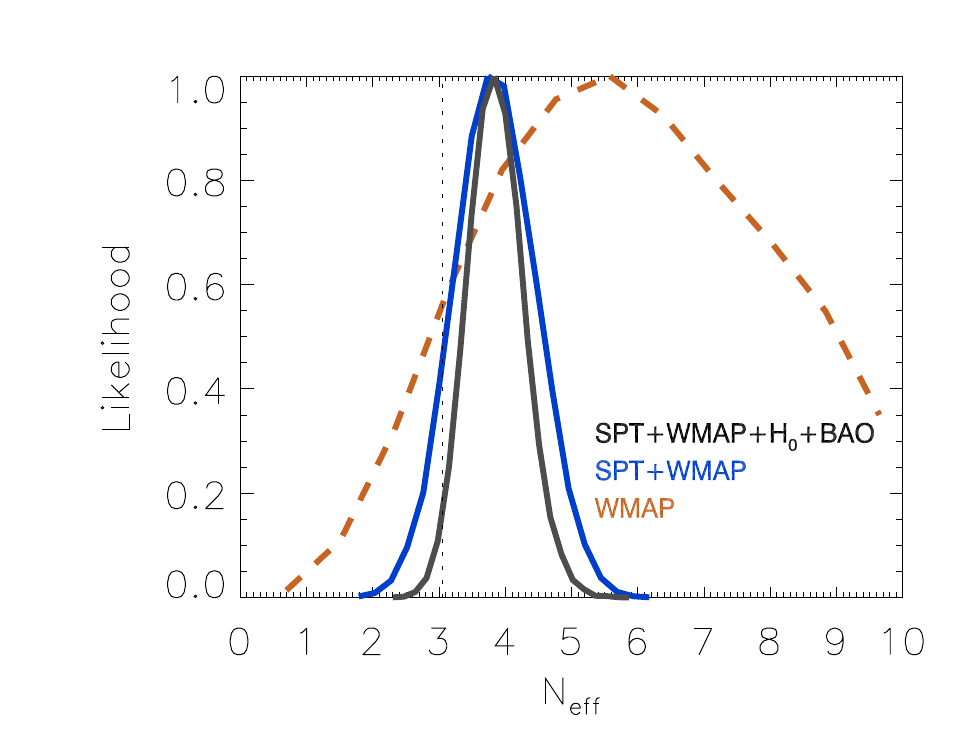}
  \caption[]{ The one-dimensional marginalized constraint on the effective number of relativistic species \neff.  The standard value of $\neff=3.046$ is shown by the vertical dotted line.}
  \label{fig:neff}
\end{figure*}

We can extend this test by promoting \neff\ to a free parameter.  \citet{komatsu11} find $\neff>2.7$ (95\% CL) using WMAP7 alone, while \citet{dunkley10} find $\neff=5.3 \pm 1.3$ using ACT+WMAP7.  The SPT+WMAP7 data constrain \neff~to be 

\begin{equation}
\neff = 3.85 \pm 0.62.
\end{equation}
This constraint is $1.3\sigma$ higher than the standard $\neff=3.046$.  When the $H_0$ and BAO data are added, the constraint improves to $\neff=3.86 \pm 0.42$, 1.9$\sigma$ higher than the standard $\neff=3.046$.  As discussed in Section~\ref{sec:discuss_damping} and shown in Table~\ref{tab:params_hbao_clusters}, the constraint is $\neff=3.42 \pm 0.32$, 1.2$\sigma$ higher than the standard value, when information from local galaxy clusters is added.  Figure~\ref{fig:neff} shows the one-dimensional marginalized constraint on \neff.

\neff\ has also been constrained using measurements of abundances of $^4$He and deuterium.  These abundances are sensitive to the expansion rate during BBN, which, in turn, is sensitive to \neff.  Using these methods, \citet{simha08} find $\neff=2.4 \pm 0.4$, although the $^{4}$He abundance used in that work, $\yhe=0.240 \pm 0.006$, is lower than more recent determinations (\eg\ $\yhe=0.2609 \pm 0.0117$ from \citet{aver11}).  \citet{mangano2011} use $^4$He and deuterium abundances to provide a conservative upper limit of \neff$<4.2$ (95\% CL).

The effective number of relativistic species has also been constrained by combining low-redshift measurements with WMAP5 data.  For example, \citet{reid10} found $\neff=3.76^{+0.63}_{-0.68}$ using the abundance of optically-selected galaxy clusters, CMB data, and a measurement of $H_0$.  Similarly, \citet{mantz10c} found $\neff=3.4^{+0.6}_{-0.5}$ using the abundance of X-ray-selected galaxy clusters, galaxy cluster gas mass fraction data, CMB data, supernova data, BAO data, and a measurement of $H_0$.  These results are consistent with, but not independent from, the constraints presented here, as WMAP data is common to all of these constraints.

\begin{table*}[h]
\begin{center}
\begin{threeparttable}
\caption[Constraints on Cosmological Parameters using SPT+WMAP+$H_0$+BAO+Clusters]{Constraints on Cosmological Parameters using SPT+WMAP+$H_0$+BAO+Clusters}
\begin{tabular}{ | l  c  c  c  c | }
\hline \hline
 & & $\Lambda$CDM & $\Lambda$CDM & $\Lambda$CDM \\
 & & +~\nrun & +~\yhe & +~\neff \\
  \hline \hline
Primary & $100\Omega_{b}h^2$ & $2.23\pm0.040$ & $2.26\pm0.045$ & $2.24\pm0.041$\\
Parameters & $\Omega_{c}h^2$ & $0.111\pm0.0020$ & $0.111\pm0.0020$ & $0.116\pm0.0054$\\
 & $100\theta_s$ & $1.041\pm0.0016$ & $1.043\pm0.0019$ & $1.040\pm0.0017$\\
 & \ns & $0.9751\pm0.0110$ & $0.9787\pm0.0123$ & $0.9757\pm0.0116$\\
 & $\tau$ & $0.0897\pm0.015$ & $0.0852\pm0.014$ & $0.0821\pm0.014$\\
 & $10^9 \Delta^2_{R}$ & $2.33\pm0.092$ & $2.35\pm0.082$ & $2.37\pm0.081$\\
\hline
Extension & \nrun & $-0.017\pm0.012$ & --- & ---\\
Parameters & \yhe & ($0.2478\pm0.0002$) & $0.288\pm0.029$ & ($0.2526\pm0.004$)\\
 & \neff & (3.046) & (3.046) & $3.42\pm0.32$\\
\hline
Derived & $\sigma_8$ & ($0.809\pm0.014$) & ($0.819\pm0.016$) & ($0.823\pm0.019$)\\
\hline
 & $\chi^2_{\rm min}$ & 7509.3 & 7509.3 & 7510.3\\
  \hline \hline
\end{tabular}
\label{tab:params_hbao_clusters}
\begin{tablenotes}
\item The constraints on cosmological parameters using SPT+WMAP7+$H_0$+BAO+Clusters, where ``Clusters'' refers to the local cluster abundance measurement of \citet{vikhlinin09}.  We report the mean of the likelihood distribution and the symmetric 68\% confidence interval about the mean.  The label ``---'' signifies $\nrun=0$.
\end{tablenotes}
\end{threeparttable}
\end{center}
\end{table*}

\subsection{Discussion of Models that allow for Additional Damping}
\label{sec:discuss_damping}

In the previous sections we have fit the SPT+WMAP data (and in some cases, $H_0$ and BAO data) to a flat, $\Lambda$CDM cosmological model and to a number of extensions beyond this model.  Of these extensions, three of them---allowing for spectral running, varying the primordial helium abundance, and varying the effective number of relativistic species---improved the fit to the data by $\Delta\chi^2 \sim 3$, primarily by lowering the predicted power in the CMB damping tail.  In other words, models with either a negative spectral running, a high helium abundance, or a high effective number of relativistic species were mildly preferred, at 1.6-1.9$\sigma$ depending on the model.  We use this section to briefly discuss the robustness of the preference, the degeneracies between the parameters in these models, and the consistency (or lack thereof) of these models with external datasets.  We find that these models prefer values of $\sigma_8$ that are disfavored by measurements of local galaxy clusters, and that including the cluster information brings the constraints on these parameters closer to their standard values.

\subsubsection{Robustness of Preference}
\label{sec:robust_pref}
The models with spectral running and free \yhe\ prefer non-standard models  at 1.8$\sigma$ and 1.6$\sigma$, respectively, using only the CMB data, and these preferences shift to 1.7$\sigma$ if the $H_0$ and BAO data are included.  The model with free \neff\ deviates from the standard $\neff=3.046$ by 1.3$\sigma$ when only CMB data are used, and a high \neff\ is preferred at 1.9$\sigma$ only after the $H_0$ and BAO data are combined with the CMB data.  The data do not show any significant preference for one extension over the others.

We have tried doubling the SPT beam uncertainty, doubling the widths of the priors on the foreground parameters, and using a ``$D_\ell^{\rm CL}=\rm{constant}$'' template for the clustered point source power, and find that none of these significantly weakens the preference for additional damping.

\subsubsection{Degeneracies between Parameters}
\label{sec:degen}
The best-fit model spectrum is lower by $\sim$2.5\% at $\ell=2000$ for each of the three  models relative to the baseline model.  This suggests that the extension parameters \nrun, \yhe, and \neff\ are degenerate.  Indeed, we find that for models in which two or three of these parameters are free, there are degeneracies between the extension parameters, and the constraint on any one parameter is weakened.

\begin{figure*}[h]\centering
\includegraphics[]{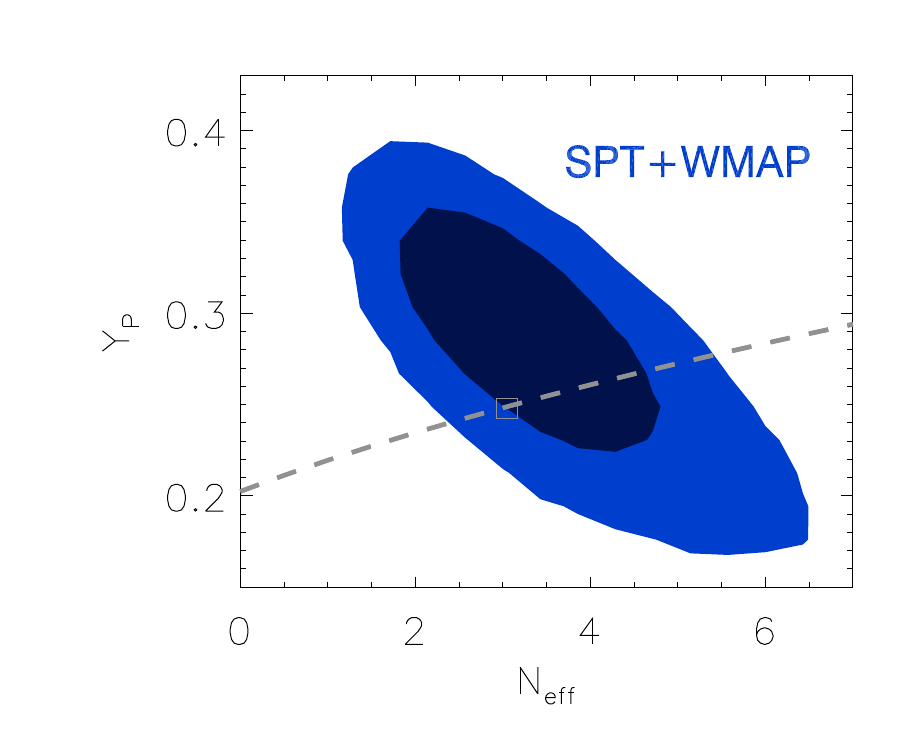}
  \caption[]{ The two-dimensional marginalized constraint on the primordial helium abundance \yhe\ and the effective number of relativistic species \neff\ for a model in which both parameters are free.  The two-dimensional contours show the 68\% and 95\% confidence intervals.  The relation between the two quantities in standard BBN theory is shown by the dashed line, with the point ($\neff=3.046$, $\yhe=0.2478$) shown by the square.  The constraint on \neff\ shown in Figure~\ref{fig:neff} is essentially a cut through this likelihood along the BBN curve, while the constraint on \yhe\ shown in Figure~\ref{fig:yhe} is a cut along $\neff=3.046$.}
  \label{fig:yhe_neff}
\end{figure*}

To illustrate this degeneracy, consider a model in which the primordial helium abundance \yhe\ and the effective number of relativistic species \neff\ are both free.  In this model, the two are no longer related by BBN theory and are independent.  In Figure~\ref{fig:yhe_neff}, we show the two-dimensional constraint on these parameters using SPT+WMAP7.  First, note that there is a degeneracy between the parameters; a cosmology with a high \neff\ can be accommodated by lowering \yhe.  But the two are not completely degenerate, as the contours do not extend to \neff=0 or \yhe=0.  The marginalized constraints are $\neff=3.4 \pm 1.0$ and $\yhe=0.283 \pm 0.045$.  This is an interesting result in itself: the SPT+WMAP7 data are able to significantly detect the effects of helium and neutrinos independently.  The main point, however, is that the three extension parameters that affect damping-tail power---\nrun, \yhe, and \neff---are degenerate, and the constraint on any one of them is reduced if we allow the others to be free.  For simplicity we have presented the results from models where only one of these parameters is free, rather than from the models where two or more of them are simultaneously free.

\subsubsection{Consistency with External Data}
\label{sec:external}
Are any of these scenarios---a negative spectral running, a high \yhe, or a high \neff---obviously ruled out by existing data?  The spectral running is primarily constrained through the CMB, and previous measurements do not rule out the SPT+WMAP7 best-fit value, $\nrun= -0.024 \pm 0.013$.  Regarding helium, the SPT+WMAP7 best-fit value of $\yhe= 0.296 \pm 0.030$ is consistent at 1.2$\sigma$ with the HII-region-based measurements of \citet{aver11}, who find $\yhe=0.2609 \pm 0.0117$.  Regarding the number of relativistic species, the SPT+WMAP7+$H_0$+BAO best-fit value for the effective number of relativistic species, $\neff=3.86 \pm 0.42$, is consistent with the upper limit of $\neff<4.2$ (95\% CL) inferred by \citet{mangano2011} using $^4$He and deuterium abundances.  It also consistent with, but not independent from, the results of \citet{reid10} and \citet{mantz10c}, which combine WMAP5 data with low-redshift measurements.

A common feature of these three models is that they require values of $\sigma_8$, the amplitude of linear matter fluctuations on scales of $8~h^{-1}$ Mpc at $z=0$, that are higher than those favored in the baseline model.  The $\sigma_8$ required in the high-\neff\ model is particularly high: the constraint on $\sigma_8$ is $\sigma_8 = 0.871 \pm 0.033$ using SPT+WMAP+$H_0$+BAO, while the equivalent constraint for the baseline model is $\sigma_8 = 0.818 \pm 0.019$.  This correlation between high-\neff\ models and high $\sigma_8$ was noted in \citet{dunkley10}.  Are such high values of $\sigma_8$ consistent with low-redshift measurements?  To answer this question, we consider the galaxy cluster abundance measurement of \citet{vikhlinin09}, which directly and tightly constrains $\sigma_8$ and is consistent with other measurements of structure at low to medium redshift \citep{rozo10, mantz10b, vanderlinde10, sehgal10b}.  The authors use the abundance of local ($0.025<z<0.22$) clusters to infer $\sigma_8 (\Omega_m/0.25)^{0.47} = 0.813\pm 0.013\pm 0.024$, where the second set of errors is an estimate of the systematic uncertainty due to the uncertainty in the masses of the clusters.  This result is essentially independent of the data we have used thus far and is not affected by varying \neff.  

\begin{figure*}[h]\centering
\includegraphics[width=1.0\textwidth]{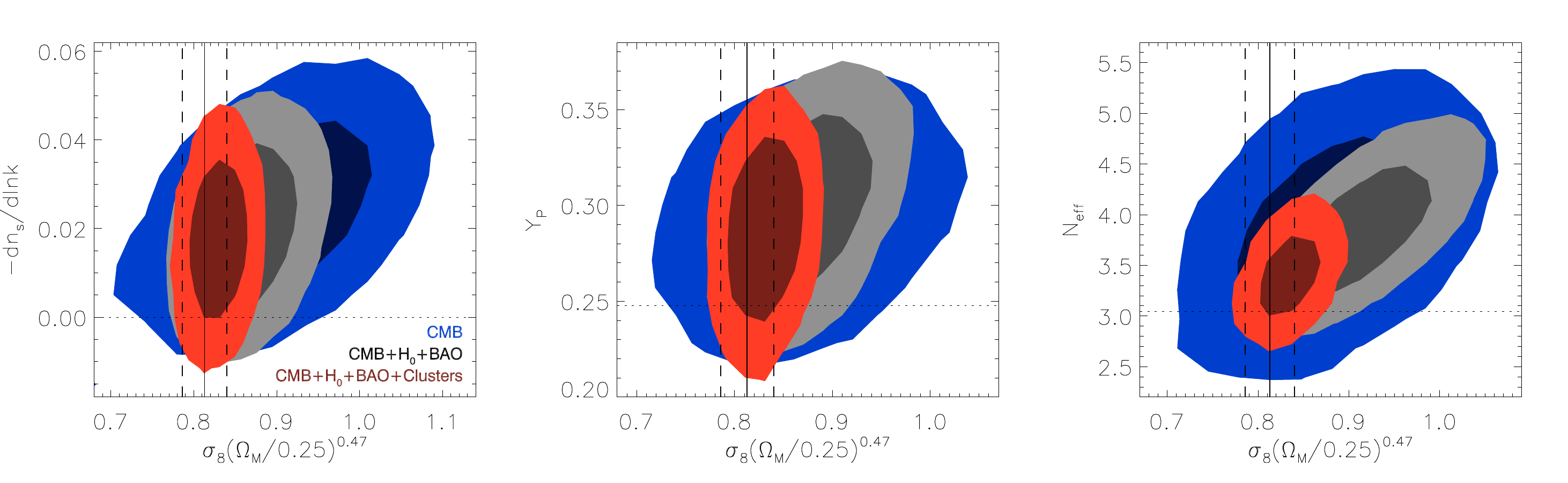}
  \caption[]{ The two-dimensional marginalized constraints on spectral running, primordial helium, or the effective number of relativistic species versus the combination $\sigma_8(\Omega_M/0.25)^{0.47}$, which is well constrained by the cluster abundance measurement of \citet{vikhlinin09}.  Each panel corresponds to a distinct Markov chain.  ``CMB'' corresponds to SPT+WMAP7.  The two-dimensional contours show the 68\% and 95\% confidence intervals.  The constraint on $\sigma_8(\Omega_M/0.25)^{0.47}$ from the clusters and the corresponding 1$\sigma$ uncertainties are shown by the vertical lines.  The standard values of the spectral running, primordial helium, and the effective number of relativistic species are shown by the dotted horizontal lines.  Adding the cluster abundance information moves the constraints on these parameters closer to their standard values.}
  \label{fig:cluster}
\end{figure*}

As can be seen in Figure~\ref{fig:cluster}, the cluster data prefer a value of $\sigma_8(\Omega_m/0.25)^{0.47}$ that is lower than those preferred by SPT+WMAP+$H_0$+BAO in the \nrun, free-\yhe, or free-\neff\ models.  Put another way, the constraints on these parameters move closer to their standard values when the cluster information is included.  This effect is most significant for the constraint on the effective number of relativistic species, which moves to $\neff= 3.42 \pm 0.32$ and is 1.2$\sigma$ from the standard value of $\neff=3.046$.  Similarly, the spectral running and primordial helium abundance are moved to within 1.4$\sigma$ of their standard values when the cluster information is included.  We conclude that models with a negative spectral running, a high \yhe, or a high \neff\ are disfavored by the cluster abundance data.  The full parameter constraints for these models using SPT+WMAP+$H_0$+BAO+Clusters are given in Table~\ref{tab:params_hbao_clusters}.

We also note that lower values of $\sigma_8$ are obtained if neutrinos are allowed to have mass.  For example, for a model in which $\neff$ and $\sum m_\nu$ are allowed to be free, we find $\neff = 3.98 \pm 0.43$, $\sigma_8 = 0.803 \pm 0.056$, and $\sum m_\nu < 0.69$ eV (95\% U.L.) using SPT+WMAP+$H_0$+BAO (compared to $\neff=3.86 \pm 0.42$ and $\sigma_8 = 0.871 \pm 0.033$ if neutrinos are forced to be massless).

\section{Conclusion}
\label{sec:conclusion}

We have presented a new measurement of the damping tail of the CMB power spectrum using data from the South Pole Telescope.  This measurement builds upon earlier measurements of the damping tail by ACBAR \citep{reichardt09a}, QUaD \citep{brown09, friedman09} and ACT \citep{das10}.  The SPT power spectrum uses $150$~GHz data and spans the multipole range $650 < \ell < 3000$, where it is dominated by primary CMB anisotropy.  We combine this spectrum with data from WMAP7 to constrain cosmological models.  We find that the SPT and WMAP7 spectra are consistent with each other, and that when combined they are well fit by a spatially flat, $\Lambda$CDM cosmology.

The addition of the SPT data provides modest improvements to the constraints on the standard six-parameter model relative to using WMAP alone.  One notable improvement is that SPT+WMAP7 measure the scalar spectral index to be $\ns=0.9663 \pm 0.0112$, which disfavors the Harrison-Zel'dovich-Peebles index ($\ns=1$) at 3.0$\sigma$ using only CMB data.  When low-redshift measurements of the Hubble constant \citep{riess11} and the BAO feature \citep{percival10} are included, the constraint on the scalar spectral index improves to $n_s = 0.9668\pm0.0093$, a 3.6$\sigma$ rejection of $n_s=1$.

We consider a number of extensions beyond this baseline model.  First we consider a model in which the amplitude of gravitational lensing on the CMB is allowed to vary freely, and find that the SPT+WMAP data detect, at $\sim$5$\sigma$, the effect of gravitational lensing, and that the amplitude is consistent with the $\Lambda$CDM cosmological model.  Parametrized in terms of a rescaling of the lensing potential power spectrum ($C_\ell^{\phi\phi} \rightarrow A_{L} C_\ell^{\phi\phi}$), the lensing amplitude is $A^{0.65}_{L}=0.94 \pm 0.15$.

We consider a model in which the power from tensor fluctuations is allowed to vary freely.  We constrain the tensor-to-scalar ratio to be $r<0.21$ (95\% CL) using SPT+WMAP7, and $r<0.17$ (95\% CL) using SPT+WMAP7+$H_0$+BAO.

We consider a model in which the scalar spectral index $n_s$ is allowed to vary or ``run'' as  function of wavenumber.  We constrain the spectral running to be $\nrun= -0.024 \pm 0.013$ using SPT+WMAP7.

We consider a model in which the primordial helium abundance, typically a function of standard BBN theory, is allowed to vary freely.  That is, we measure the effect of helium due solely to its effect on the CMB damping tail.  We strongly detect the effect of helium on the CMB; a model with no helium is rejected at 7.7$\sigma$.  When the primordial helium abundance is allowed to vary freely, we find $\yhe=0.296 \pm 0.030$ using SPT+WMAP7.

Finally, we consider a model in which the effective number of relativistic species in the early universe is allowed to vary freely.  Normally this is the number of neutrinos, three, plus a small correction due to electron-positron energy injection, resulting in $\neff^\mathrm {standard}=3.046$.  Using SPT+WMAP7 we strongly detect the effect of neutrinos on the CMB; a model with no neutrinos is rejected at 7.5$\sigma$.  When \neff\ is allowed to vary freely, we find $\neff=3.85 \pm 0.62$, while using SPT+WMAP7+$H_0$+BAO we find $\neff=3.86 \pm 0.42$.

Three of these model extensions---spectral running, free helium, and free \neff---show a mild, $\sim$1.7$\sigma$ preference for non-standard models.  We find that such models are disfavored by the value of $\sigma_8$ inferred from the abundance of low-redshift galaxy clusters \citep{vikhlinin09}.  The constraints on these parameters move closer to their standard values when the cluster information is included.  Using SPT+WMAP7+$H_0$+BAO+Clusters, the constraints are $\nrun= -0.017 \pm 0.012$, $\yhe=0.288 \pm 0.029 $, and $\neff=3.42 \pm 0.32$.

The SPT data presented here cover 790 square degrees.  The full SPT-SZ survey, which is expected to be completed by the end of 2011, will cover approximately 2500 square degrees.   With 150 GHz data of the quality used here and with additional data at 90 and 220 GHz, a power spectrum analysis of the full SPT survey should be at least 1.7 times more sensitive than that presented here.

\begin{acknowledgments}
The South Pole Telescope is supported by the National Science
Foundation through grants ANT-0638937 and ANT-0130612.  Partial
support is also provided by the NSF Physics Frontier Center grant
PHY-0114422 to the Kavli Institute of Cosmological Physics at the
University of Chicago, the Kavli Foundation and the Gordon and Betty
Moore Foundation.  
The McGill group acknowledges funding from the National   
Sciences and Engineering Research Council of Canada, 
Canada Research Chairs program, and 
the Canadian Institute for Advanced Research. 
We acknowledge use of the FNAL-KICP Joint Cluster.
R. Keisler acknowledges support from NASA Hubble Fellowship grant HF-51275.01.
B.A. Benson is supported by a KICP Fellowship.
M. Dobbs acknowledges support from an Alfred P. Sloan Research Fellowship.
L. Shaw acknowledges the support of Yale University and NSF grant AST-1009811.  
M. Millea and L. Knox acknowledge the support of NSF grant 0709498.  
This research used resources of the National Energy Research Scientific Computing Center, which is supported by the Office of Science of the U.S. Department of Energy under Contract No. DE-AC02-05CH11231. 
We acknowledge the use of the Legacy Archive for Microwave Background Data Analysis (LAMBDA). 
Support for LAMBDA is provided by the NASA Office of Space Science.
\end{acknowledgments}

\clearpage

\bibliography{../../BIBTEX/spt.bib}

\end{document}